\documentclass[
    ]
    {aa}

\usepackage{graphicx}
\graphicspath{{figures/}}
\usepackage[varg]{txfonts}  

\usepackage{color,graphicx}
\usepackage{amsfonts}
\usepackage{amssymb}
\usepackage{graphicx}
\usepackage{epsfig}
\usepackage{tikz}

\usepackage{amssymb}
\usepackage{amscd}
\usepackage{amsmath} 
\usepackage{mathrsfs}
\usepackage{xspace}
\usepackage{mathrsfs}
\usepackage{dsfont}
\usepackage{bbm}
\usepackage{slashed}

\usetikzlibrary{decorations.pathmorphing,patterns}

\usepackage{commath}
\usepackage{siunitx}[=v2]
\usepackage[super]{nth}

\usepackage{xcolor}

\usepackage{hyperref}
\hypersetup{
    pdffitwindow=true,
    pdfstartview={FitH},
    pdftitle={Waterfall PCA Folding},
    pdfauthor={Giampiero Naletto, Tomas Cassanelli},
    pdfcreator={Tomas Cassanelli},
    pdfproducer={Giampiero Naletto, Tomas Cassanelli},
    pdfnewwindow=true,
    colorlinks=true,
    linkcolor=blue,
    citecolor=blue,
    urlcolor=blue
    }

\begin{document}

\title{Accretion flows around exotic tidal wormholes}
\subtitle{I. Ray-tracing}
\author{
	O.\ Sokoliuk\inst{\ref{1},\ref{2}}
	\and S. Praharaj \inst{\ref{3}}
	\and A.\ Baransky\inst{\ref{2}}
	\and P.K. Sahoo\inst{\ref{3}}
	}
\institute{
	Astronomical Observatory of the National Academy of Sciences of Ukraine (MAO NASU), Kyiv, 03143, Ukraine. \label{1} \\
	\email{oleksii.sokoliuk@mao.kiev.ua} 
	\and
	Astronomical Observatory, Taras Shevchenko National University of Kyiv, 3 Observatorna St., 04053 Kyiv, Ukraine. \label{2} \\
	\email{abaransky@ukr.net}
	\and
	Department of Mathematics, Birla Institute of Technology and
    Science-Pilani, Hyderabad Campus, Hyderabad-500078, India \label{3}\\
    \email{pksahoo@hyderabad.bits-pilani.ac.in, f20180714@hyderabad.bits-pilani.ac.in}
	}


	\abstract
	{}
	{This paper investigates the various spherically symmetric wormhole solutions in the presence of tidal forces and applies numerous methods, such as test particle orbital dynamics, ray-tracing and microlensing.}
	{We make the theoretical predictions on the test particle orbital motion around the tidal wormholes with the use of normalized by $\mathcal{L}^2$ effective potential. In order to obtain the ray-tracing images (of both geometrically thin and thick accretion disks, relativistic jets), we properly modify the open source \texttt{GYOTO} code with python interface}
	{We applied this techniques to probe the accretion flows nearby the Schwarzschild-like and charged Reissner-N\"ordstrom (RS) wormholes (we assumed both charged RS wormhole and special case with the vanishing electromagnetic charge, namely Damour-Solodukhin (DS) wormhole). It was shown that the photon sphere for Schwarzschild-like wormhole presents for both thin and thick accretion disks and even for the vanishing tidal forces. Moreover, it was observed that $r_{\mathrm{ph}}\to\infty$ as $\alpha\to\infty$, which constraints $\alpha$ parameter to be sufficiently small and positive in order to respect the EHT observations. On the other hand, for the case of RS wormhole, photon sphere radius shrinks as $\Lambda\to\infty$, as it was predicted by the effective potential. In addition to the accretion disks, we as well probe the relativistic jets around two wormhole solutions of our consideration. Finally, with the help of star bulb microlensing, we approximate the radius of the wormhole shadow and as we found out, for Schild WH, $R_{\mathrm{Sh}}\approx r_0$ for ZTF and grows linearly with $\alpha$. On the contrary, shadow radius for charged wormholes slowly decreases with the growing trend of DS parameter $\Lambda$.}
	{}


	\maketitle

\section{Introduction}\label{sec:1}
It is generally believed that such massive and compact objects as supermassive black holes, located within the central galactic regions gain their mass mainly from the accreting processes. Moreover, in order to keep relatively high accretion rates, around such massive objects the so-called accretion disks exists. They are formed by a rotating fluid that slowly spirals up to the surface of massive object (in the case of black holes, up to the event horizon) \citep{Liu:2021yev}. Part of the heat that magnetised fluid produces is converted into the radiation that could be observed in both optical and X-ray ranges. Moreover, existence of such objects as accretion disks were confirmed with the use of various methods, such as VLBI direct imaging \citep{EventHorizonTelescope:2019dse,EventHorizonTelescope:2019uob,EventHorizonTelescope:2019uob,2019ApJ...875L...4E,EventHorizonTelescope:2019pgp,EventHorizonTelescope:2019ggy}, LIGO and VIRGO observations \citep{PhysRevLett.116.061102,PhysRevLett.116.221101}, observations of the ultra-high resolution electromagnetic spectrum \citep{doi:10.1146/annurev-astro-082812-141003,Nampalliwar_2020}. 

The first complete model for geometrically thin accretion disk was proposed in the pioneering work of Shakura and Sunyaev \citep{1973A&A....24..337S}. In the following decades, it was consequently modified for the use in the general theory of relativity by Novikov \& Thorne \citep{1973blho.conf..343N}, Page \& Thorne \citep{1974ApJ...191..499P} respectively. Aforementioned models in the present days are being applied for the detailed investigations of various black hole spacetimes \citep{Lin:2022ksb,Mirzaev:2022xpz,Wang:2019tjc} and even more exotic objects, such as wormhole-like solutions \citep{RAHAMAN2021115548,Paul_2020}. As a massive compact object in the current work we consider two exotic models of wormholes.
\subsection{Traversable wormholes}
Apart from the black holes, there exist another class of compact objects, namely wormhole-like objects. Cosmological wormholes could be interpreted as a two-way connections of a far regions within a four-dimensional manifold. Besides, they could connect two distinct universes or wormhole throat could even be laid through the higher dimensional bulk. Obviously, such solutions does not have event horizon, and therefore if two-sided accretion disk is present, wormhole solution could be easily distinguished from the black hole. However, it was shown that wormholes could mimic the behavior of black holes and many of their properties \citep{Cardoso2019TestingTN}. There was a lot of propositions for the wormhole solutions during the last few decades. First ever WH-like solution were introduced in the work of Flamm \citep{Flamm2015} and then was modified by the Einstein \& Rosen \citep{PhysRev.48.73}, was called an Einstein-Rosen bridge. It was found that ER bridge cannot be humanly traversable, since at the ER wormhole throat a singularity is present \citep{PhysRev.128.919}. First interesting model of traversable wormhole was presented only in the 1988 by Morris and Thorne \citep{1988AmJPh..56..395M}. Unfortunately, it was discovered that in order for Morris-Thorne wormhole to be traversable, the so-called exotic matter needs to be present at wormhole throat (exotic matter does violate Null Energy Condition (NEC) $T_{\mu\nu}k^\mu k^\nu\geq0$ that comes from the Raychaudhuri equation). For the sake of NEC validation, there was many various methods applied, such as assumption of gravity modification \citep{PhysRevD.87.067504}, presence of the additional matter fields \citep{https://doi.org/10.48550/arxiv.gr-qc/0201083,Caceres:2019giy} or creation of the wormhole system, where quantum effect competes with the classical ones \citep{Visser:1995cc,Gao:2016bin,https://doi.org/10.48550/arxiv.1804.00491}. In general, despite of the many problems present within the field of traversable wormholes, there is growing interest in them present, since traversable wormholes could provide a non-singular replacement of black holes in the general theory of relativity.

In the works of \citep{Bhattacharyya2001,TORRES2002377,Yuan_2004,PhysRevD.73.021501}, emissivity of accretion disks around different exotic objects, such as quark, fermion and boson stars (both rotating and non-rotating) were investigated. Moreover, there is some studies present on the topic of thin accretion disks around naked, Bogush-Galtsov and strongly naked singularities (for more details on the subject, see papers \citep{PhysRevD.82.124047,PhysRevD.100.024055,Gyulchev2020,Karimov:2022qdb}). Finally, observational signatures of the rotating and non-rotating wormholes were examined in \citep{RAHAMAN2021115548,Paul_2020}, of Ellis-Bronnikov wormhole in \citep{Yusupova:2021rcz}. There was also some comparative studies present, that for example compare thin accretion disk images, obtained for rotating gravastars and Kerr black holes \citep{Harko:2009gc}. Moreover, it was numerically shown in \citep{Karimov:2019qfw} that Damour-Solodukhin wormholes are practically could not be distinguished from the Kerr back holes in terms of accretion. Finally, in \citep{Karimov:2020fuj} both naked singularity, wormhole and black hole were compared in the sense of accretion properties. Aforementioned studies could help us to discriminate black holes from other, more exotic objects of non-singular nature.

In the present article, we are going to focus our investigation on the two special forms of Morris-Thorne wormhole, namely Schwarzschild-like wormhole and Reissner-N\"ordstrom wormhole (generalised case of Damour-Solodukhin wormhole with the present electromagnetic charge). We are going to study both geometrically thin and thick accretion disks with the use of numerical ray tracing methods and effective potential.
\subsection{Article organisation}
In the current subsection, we will briefly present the organisation of our paper. In the Section (\ref{sec:1}) we provide an introduction into the subject of thin, thick accretion disks, the history of traversable wormholes discovery and the current problems that is present in the field. Moreover, we give a bibliographical survey into the previous studies of accretion disks around various singular and non-singular cosmological objects. Consequently, in the Section (\ref{sec:2}) we introduce the Morris-Thorne wormhole geometry, conditions of wormhole traversability and two exotic wormhole models of our consideration, compute embedding surfaces for each wormhole model that we consider. In the Section (\ref{sec:3}) we hence write down the foundations of the orbital mechanics within the General Theory of Relativity and numerically compute the effective potential, it's radial derivatives for both wormhole models. In the fourth Section (\ref{sec:4}) we compute radiation flux, black body temperature and accretion disk luminosity for our WHs. Besides, in this section we produce an images of thin accreting tori around Schwarzschild and RS wormholes with the help of ray-tracing techniques. In the next Section (\ref{sec:5}) we therefore with the use of ray-tracing obtain images of thick accretion disks and blur those images up to the Event Horizon Telescope (EHT) telescope resolution of 20$\mu$as. Finally, in the Section (\ref{sec:6}) we analyse the lensing, that produce our wormhole models and in the last Section (\ref{sec:7}) we provide the concluding remarks on the key topics of our study.

\section{Exotic wormholes within GR theory}\label{sec:2}
In the present study, as we already mentioned, we are going to investigate the behavior of exotic wormhole geometries in the presence of the thin accretion disk. Generally, wormhole geometry could be reconstructed with the use of following metric tensor line element:
\begin{equation}
    ds^2=-e^{2\Phi(r)}dt^2+\bigg(1-\frac{b(r)}{r}\bigg)^{-1}dr^2+r^2d\theta^2+r^2\sin^2\theta d\phi^2
\end{equation}
Remarkably, it is easy to notice that we use mostly plus sign convention  ($\mathrm{sig}g=(-,+,+,+)$). In the line element defined above, $\Phi(r)$ is the so-called redshift function, which defines whether wormhole is tidal or not and $b(r)$ is the fundamental quantity, namely shape function, that defines wormhole geometry. For wormhole to be viable solution of field equations and satisfy traversability conditions, both shape and redshift functions must obey following statements:
\begin{itemize}
    \item $b(r_0)=r_0$, $b(r)<r$ for the case with $r>r_0$
    \item $\lim_{r\to\infty}e^{2\Phi(r)}=1$ (absence of horizons)
    \item $ \lim_{r\to\infty}b/r=0 $ (asymptotical flatness)
    \item $rb'<r$ (flaring out condition)
    \item $b'(r)\leq1$ (at least at the WH throat with $r=r_0$)
\end{itemize}
In the following, we are going to present the various exotic wormhole geometries of our consideration.

\subsection{Schwarzschild-like wormhole}
Asymptotically flat Schwarzschild-like wormhole geometries could be easily constructed with the use of shape function defined below:
\begin{equation}
    b(r)=r_0(r/r_0)^\alpha
\end{equation}
where $\alpha$ is free parameter, namely additional degree of freedom. In order to satisfy throat condition $b(r_0)/r_0=1$, condition $\alpha=(1-\beta)r_0$ must be validated, which leads to the shape function of the form \citep{2017EPJC...77..748C}:
\begin{equation}
    b(r)=(1-\beta)r_0+\beta r
\end{equation}
Case with Schwarzschild wormhole corresponds to $\beta=0$, so that mass of the wormhole reads:
\begin{equation}
    M=\frac{r_0}{2}
\end{equation}
As well, we assume that the wormhole is tidal with redshift function (for more details, see \citep{Moraes:120401,PhysRevD.57.829} and references therein):
\begin{equation}
    \Phi(r)=-\alpha/r
\end{equation}
where $\alpha>0$. In the further investigation, it will be helpful to compute the rotational velocity of the particle around each wormhole solution that we consider (following the paper \citep{Jusufi:2021lei}):
\begin{equation}
    v_{tg}^2=r\Phi'(r)=\alpha/r
    \label{eq:2.4}
\end{equation}
Now, we are going to proceed further to the next wormhole geometry that we consider.

\subsection{Reissner-Nordstr\"om black-hole-like wormhole}
Charged BH-like wormholes could be generally represented by the spherically symmetric spacetime of the form \citep{Karimov2019,PhysRevD.78.024040}:
\begin{equation}
\begin{gathered}
    ds^2=-(f(r)+\Lambda^2)dt^2+\frac{dr^2}{f(r)}
    +r^2d\theta^2+r^2\sin^2\theta d\phi^2
\end{gathered}
\end{equation}
where
\begin{equation}
    f(r)=1-\frac{2M}{r}+\frac{Q^2}{r^2}
\end{equation}
Here, $Q$ is electromagnetic charge, that lies within the bound $-M\leq Q\leq M$.
Therefore, redshift and shape functions are given as follows:
\begin{equation}
    \Phi(r)=\frac{1}{2}\log \bigg(f(r)+\Lambda^2\bigg),\quad b(r)=2M-\frac{Q^2}{r}
    \label{eq:2.11}
\end{equation}
Parameter $\Lambda$ denotes the small deviation from Reissner-N\"ordstrom black hole solution. Remarkably, even at the exponentially small values of $\Lambda$, aforementioned solution could mimic the behavior of both charged classical, semi-classical and quantum black holes \citep{PhysRevD.102.024082}. Throat of RS BH-like wormhole is located at the $r_0=M+\sqrt{M^2-Q^2}$.
As usual, conditions $\lim_{r\to\infty}b(r)=2M$ holds and therefore wormhole is massive.
Finally, we could as well derive the corresponding rotational velocity of the particle near RS BH-like WHs using equation (\ref{eq:2.11}) and (\ref{eq:2.4}):
\begin{equation}
    v_{tg}^2=\frac{M r-Q^2}{r \left(-2 M+\Lambda ^2 r+r\right)+Q^2}
\end{equation}
We already defined all of the three models of WH geometries that we are going to study in the present paper. Therefore, we could now investigate the embedding diagrams for each model.
\subsection{Embedding diagrams}
\begin{figure}
    \centering
    \includegraphics[width=0.5\textwidth]{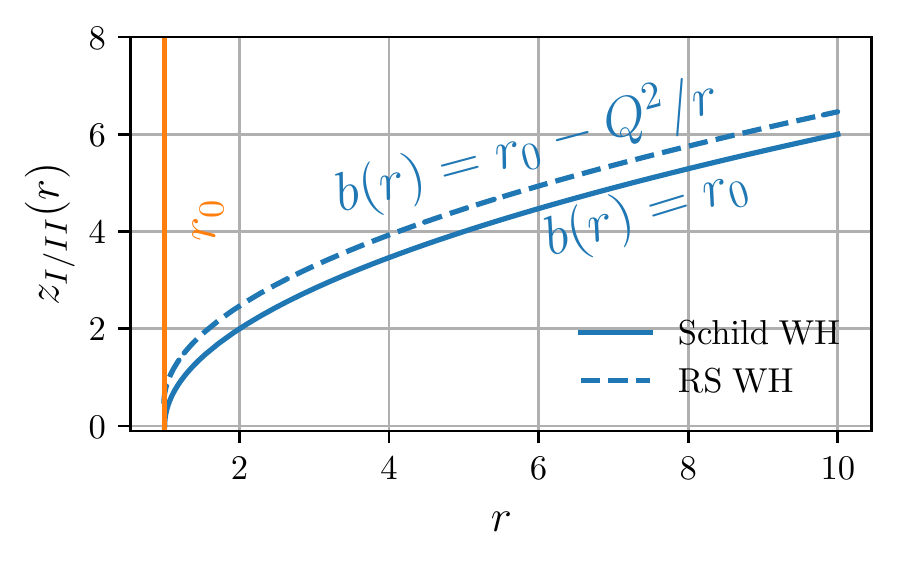}
    \caption{Wormhole embedding functions $z_I$ and $z_{II}$ for both Schwarzschild and Reissner-N\"ordstrom wormhole models with $r_0=1$, $M=0.5$ and $Q=0.1$}
    \label{fig:2}
\end{figure}

With the intention to obtain embedding diagram, we firstly need to make several assumptions. Without the loss of generality, we could assume constant time slice $t=const$ and because of the spherical symmetry we could work only in the equatorial region with $\theta=\pi/2$. With that assumptions, metric tensor becomes \citep{Jusufi2020}:
\begin{equation}
    ds^2 = \bigg(1-\frac{b(r)}{r}\bigg)^{-1}dr^2+r^2d\phi^2
    \label{eq:20}
\end{equation}
Also, it is useful to embed the reduced line element (\ref{eq:20}) to the $\mathbb{R}^3$ space (3 dimensional Euclidean space). Then, in cylindrical coordinates
\begin{equation}
    ds^2 = dz^2+dr^2+r^2d\phi^2
    \label{eq:21}
\end{equation}
Finally, by plugging the Equations (\ref{eq:20}) and (\ref{eq:21}), one could obtain embedding surface:
\begin{equation}
    \frac{dz}{dr}=\pm\sqrt{\frac{r}{r-b(r)}-1}
\end{equation}
By solving ODE above we could obtain the embedding surface:
\begin{equation}
    z_{I}=2 r_0\bigg(\sqrt{\frac{r_0}{r-r_0}}\bigg)^{-1}
\end{equation}
Embedding surface for the Reissner-N\"ordstrom wormhole could only be obtained with the use of numerical methods (in the current work, we will obtain $z_{II}$ with the help of the well-known Runge-Kutta 4th order ODE solver). Consequently, we plot the embedding diagrams for each model of our consideration with wormhole throat radius $r_0=1$ on the Figure (\ref{fig:2}). One may obviously notice that for each model, wormhole topologies are asymptotically flat, as expected. 

\section{Test particle orbital motion}\label{sec:3}
In this section, we are going to present the formalism of orbital mechanics within the General Theory of Relativity and probe the movement of test particles around exotic wormhole geometries that we consider. Because metric tensor depends only on the $r$ and $\theta$ (in our case, because of the spherical symmetry, theta is fixed to $\pi/2$, equatorial slice), there are two Killing vectors $\xi_t=\partial_t$ and $\xi_\phi=\partial_\phi$, that could be translated into conserved quantities \citep{Del_guila_2018}
\begin{equation}
    \dot{t}=\frac{\mathcal{E}}{g_{tt}}
    \label{eq:3.1}
\end{equation}
\begin{equation}
    \Ddot{r}=\frac{1}{2g_{rr}}\bigg[\partial_r g_{tt}\dot{t}^2+\partial_r g_{rr}\dot{r}^2+\partial_r g_{\phi\phi}\dot{\phi}^2\bigg]
    \label{eq:3.2}
\end{equation}
\begin{equation}
    \dot{\phi}=\frac{\mathcal{L}}{r^2}
    \label{eq:3.3}
\end{equation}
Here, as usual $\mathcal{E}$ and $\mathcal{L}$ respectively denote the conserved quantities, namely energy and angular momentum of the test particle (time- or null-like). For the timelike geodesics satisfying $ds^2=-m^2$ equation for particle's energy could be rewritten in the more suitable form:
\begin{equation}
    \mathcal{E}^2=\dot{r}^2+V_{\mathrm{eff}}\label{eq:3.4}
\end{equation}
where $V_\mathrm{eff}$ is the effective potential of gravitational field around supermassive object. For the generalized spacetime, effective potential reads \citep{Karimov2019}:
\begin{equation}
    V_\mathrm{eff}=\frac{\mathcal{E}^2g_{\phi\phi}-\mathcal{L}^2g_{tt}}{g_{tt}g_{\phi\phi}}-1
\end{equation}

\subsection{Stable circular orbits}
Now, we could discuss the different kinds of orbits that is present in orbital dynamics. Stable circular orbits in the equatorial plane could be obtained by imposing $\dot{r}=\Ddot{r}=0$ and $\dddot{r}<0$. By plugging these conditions into (\ref{eq:3.4}), we obtain that $V_{\mathrm{eff}}=dV_{\mathrm{eff}}/dr=0$ and that $d^2V_{\mathrm{eff}}/dr^2<0$. These condition implies the special values of conserved quantities, such that
\begin{equation}
    \mathcal{E}=\frac{g_{tt}}{\sqrt{g_{tt}-g_{\phi\phi}\Omega^2}}
    \label{eq:3.6}
\end{equation}
\begin{equation}
    \mathcal{L}=\frac{g_{\phi\phi}\Omega}{\sqrt{g_{tt}-g_{\phi\phi}\Omega^2}}
    \label{eq:3.7}
\end{equation}
In order to respect the equality $dV_{\mathrm{eff}}/dr$, $\Omega$ must also be constrained \citep{RAHAMAN2021115548}
\begin{equation}
    \Omega=\pm\frac{\sqrt{\partial_r g_{tt} \partial_r g_{\phi\phi}}}{\partial_rg_{\phi\phi}}
    \label{eq:3.8}
\end{equation}

\subsection{Photon sphere}
On the other hand, there are quite different requirements for the photon sphere:
\begin{equation}
    V_{\mathrm{eff}}=\mathcal{E}^2,\quad dV_{\mathrm{eff}}/dr=0,\quad d^2V_{\mathrm{eff}}/dr^2<0\label{eq:3.9}
\end{equation}
Therefore, with the help of expression for proper distance $ds^2=0$ we could get
\begin{equation}
    \mathcal{E}^2=e^{2\Phi(r)}\bigg(1-\frac{b(r)}{r}\bigg)^{-1}\dot{r}^2+V_{\mathrm{eff}}\label{eq:3.10}
\end{equation}
where for null-like geodesics, effective potential reads \citep{GODANI2021168460}:
\begin{equation}
    V_{\mathrm{eff}}=\frac{L^2e^{2\Phi(r)}}{r^2}
    \label{eq:3.11}
\end{equation}
It is interesting, that judging by the expression (\ref{eq:3.10}), wormhole throat could act as a photon sphere by itself, since $b(r_0)=r_0$, and because of this throat condition $\dot{r}=0$ automatically holds at the $r=r_0$. On the other hand, we could plot the light ray trajectories using (\ref{eq:3.2}) and (\ref{eq:3.3}):
\begin{equation}
    \bigg(\frac{dr}{d\phi}\bigg)^2=\bigg(\frac{d\tau}{d\phi}\frac{dr}{d\tau}\bigg)^2=\frac{\mathcal{L} (r-b(r)) \left(\mathcal{E}^2 r^2 e^{-2 \Phi
   (r)}-\mathcal{L}^2\right)}{r^5}
\end{equation}
Implying appropriate change of coordinates $r\to1/u$ we get
\begin{equation}
    \bigg(\frac{du}{d\phi}\bigg)^2=\left(u b\left(\frac{1}{u}\right)-1\right) \left(u^2-\frac{\mathcal{E}^2
   e^{-2 \Phi \left(\frac{1}{u}\right)}}{\mathcal{L}^2}\right)
\end{equation}
Equation above could be solved numerically in the polar coordinates $(r,\phi)$ for the special forms of both $\Phi(r)$ and $b(r)$.

\subsection{Model I}
\begin{figure*}[!htbp]
    \centering
    \includegraphics[width=\textwidth]{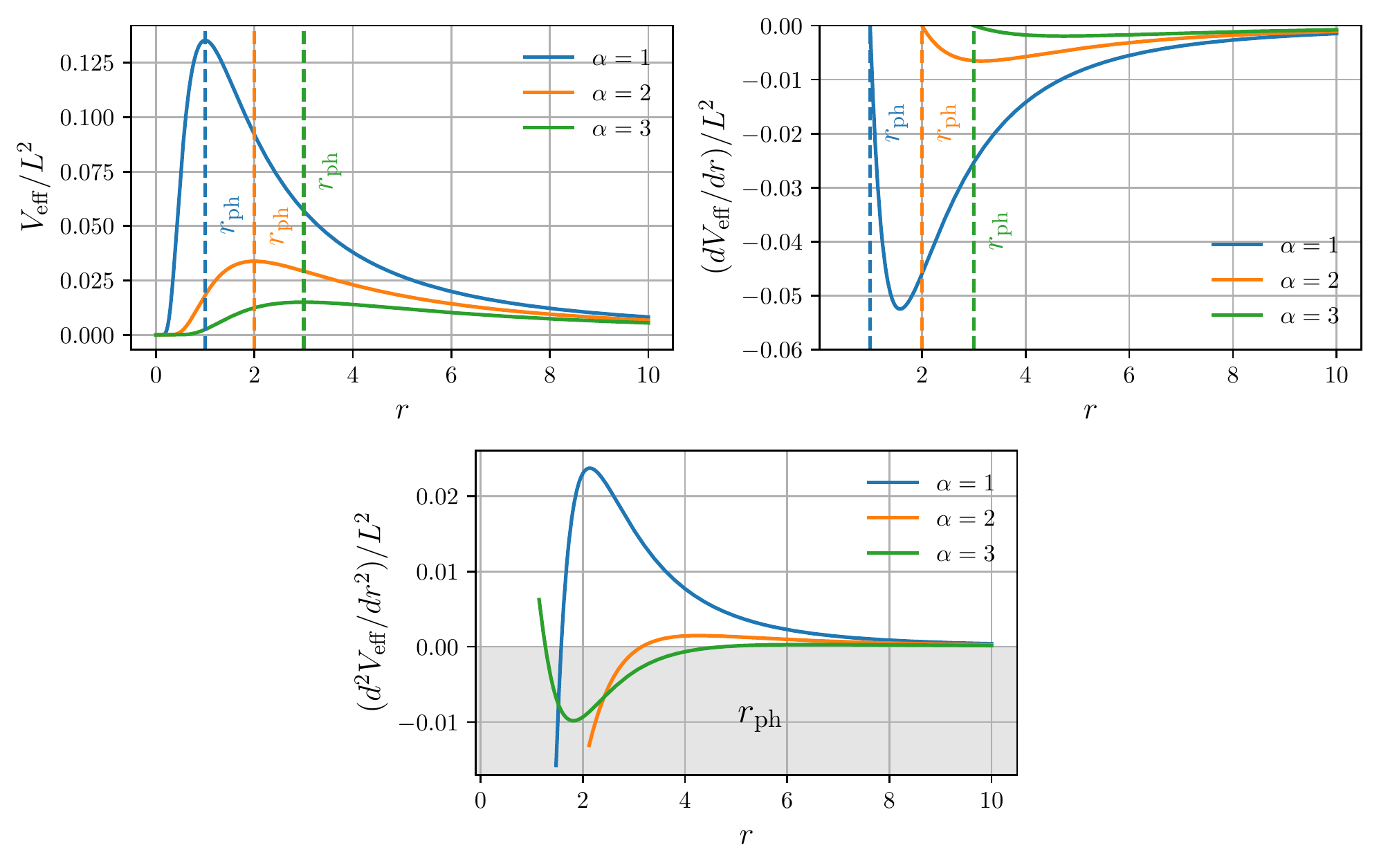}
    \caption{Normalized effective potential and it's radial derivatives for Schwarzschild-like wormhole with varying $\alpha$ parameter}
    \label{fig:3}
\end{figure*}

Now, we are going to compute the quantities (\ref{eq:3.6}), (\ref{eq:3.7}) and (\ref{eq:3.8}) for the first wormhole model, namely Schwarzschild-like wormhole with constant shape function:
\begin{equation}
    \mathcal{E}=\bigg(r \sqrt{\frac{1}{r}e^{-2 \alpha/r} (r-\alpha )}\bigg)(r-\alpha )^{-1}
\end{equation}
\begin{equation}
    \mathcal{L}=\bigg(r \sqrt{\frac{\alpha  e^{-\frac{2 \alpha
   }{r}}}{r}}\bigg)\bigg(\sqrt{\frac{e^{-\frac{2 \alpha }{r}} (r-\alpha )}{r}}\bigg)^{-1}
\end{equation}
\begin{equation}
    \Omega=\pm \frac{1}{r}\bigg(\sqrt{\frac{\alpha  e^{-\frac{2 \alpha }{r}}}{r}}\bigg)
\end{equation}
We plot normalized effective potential for Schwarzschild-like wormhole with varying free parameter $\alpha$ on the Figure (\ref{fig:3}). We marked the photon sphere on potential and it's derivatives (photon sphere is located at $r_{\mathrm{ph}}$, where effective potential has it's maximum $V_{\mathrm{eff}}(r_{\mathrm{ph}})=\mathcal{E}^2$, it's first derivative vanish and second order radial derivative is negative). As well, by solving four dimensional geodesic equations, we plot the null-geodesics with different values of impact parameter $b=|\mathcal{L}|/\mathcal{E}$ on the first plot of Figure (\ref{fig:4}).

\subsection{Model II}
\begin{figure*}[!htbp]
    \centering
    \includegraphics[width=\textwidth]{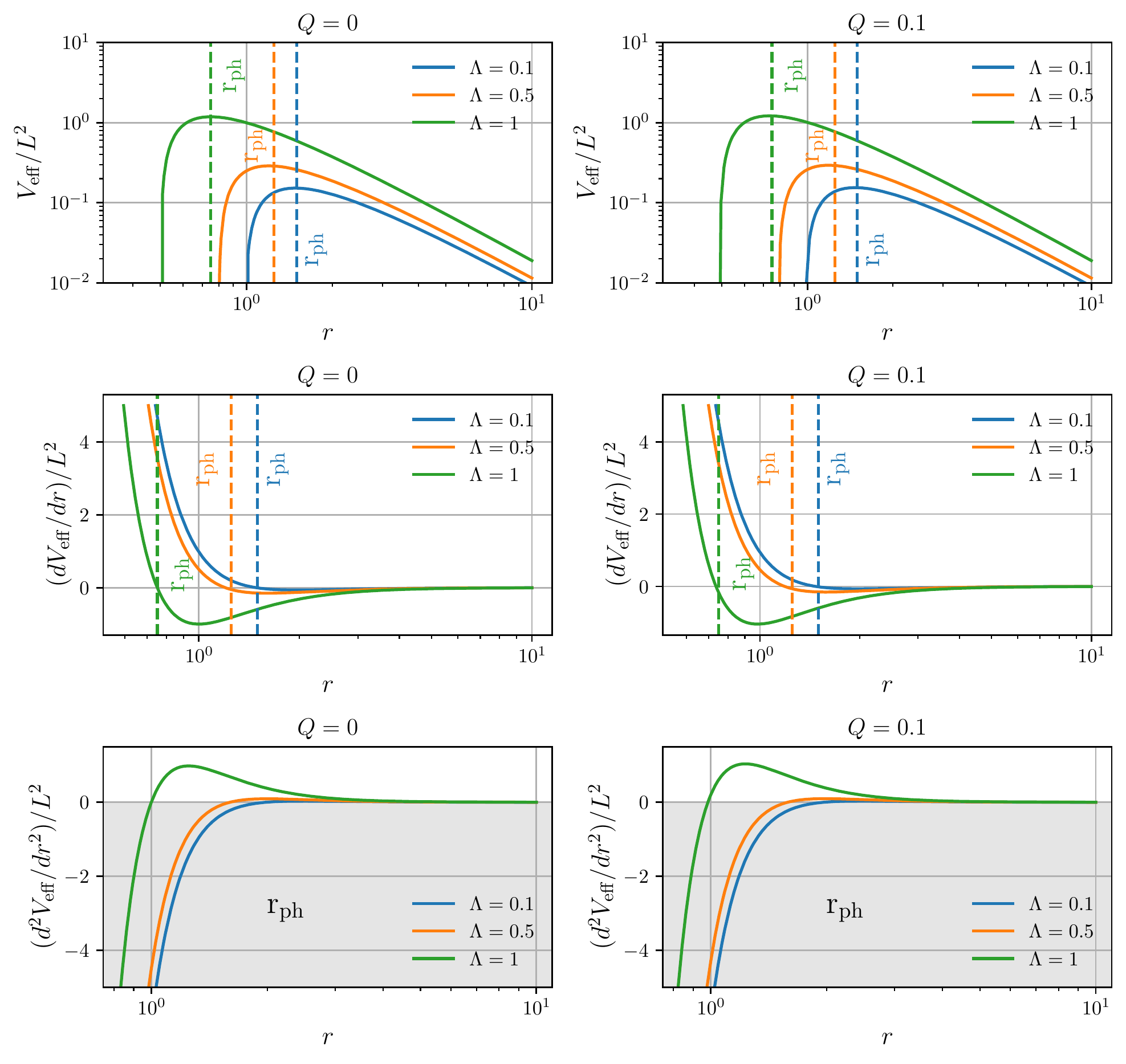}
    \caption{Normalized effective potential and it's radial derivatives for Reissner-N\"orstrom wormhole with varying electromagnetic charge $Q$ and $\Lambda$ parameter}
    \label{fig:444}
\end{figure*}

\begin{figure*}[!htbp]
    \centering
    \includegraphics[width=\textwidth]{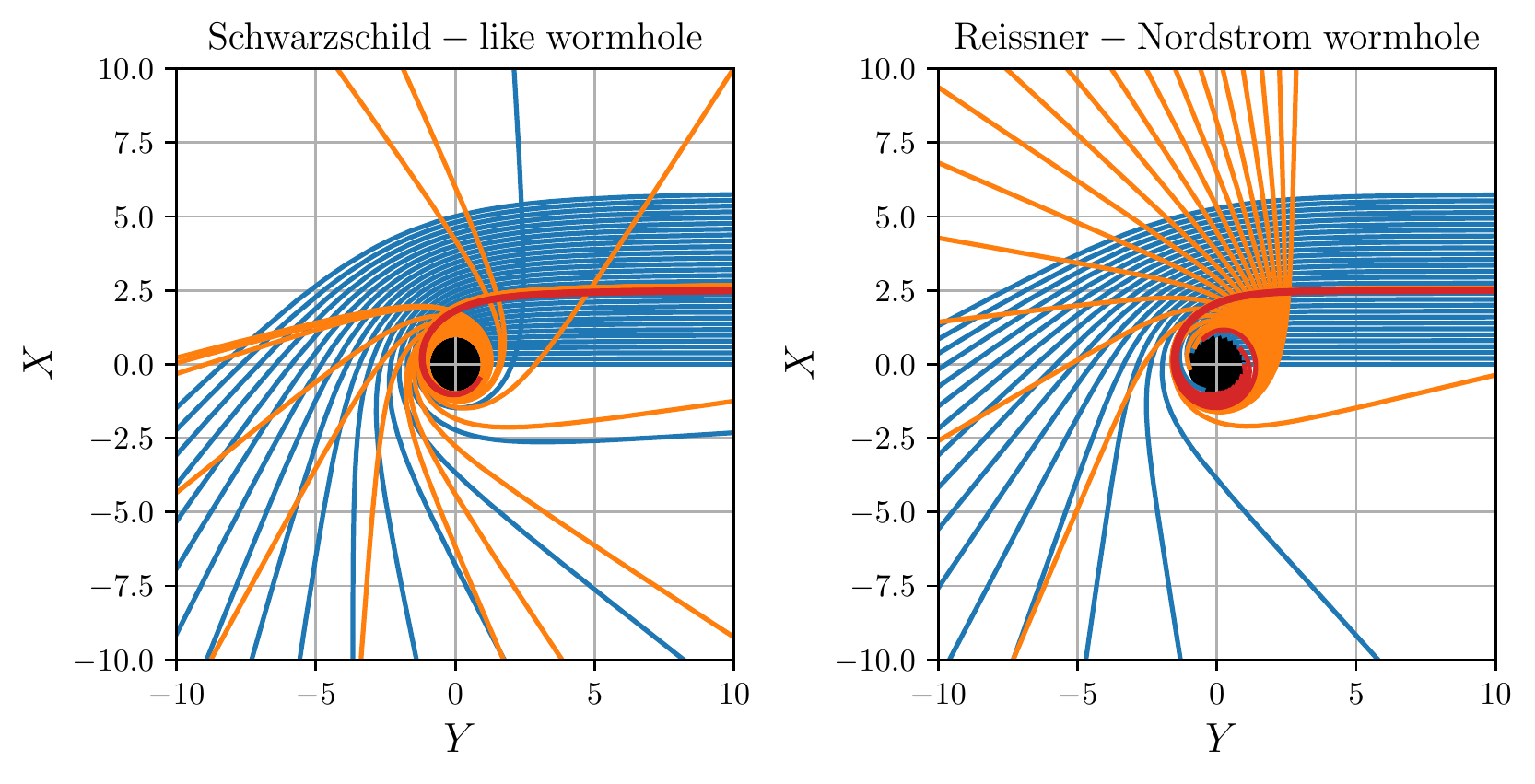}
    \caption{Orbits of null-like particles around Schwarzschild-like and Reissner-N\"ordstrom wormholes with $\alpha=2$ and $Q=\Lambda=0.1$, $M=0.5$. Here blue trajectories are direct, red - photon and orange are reconstructed with the help of lensing}
    \label{fig:4}
\end{figure*}

Following the same procedure as it was done for the Schwarzschild-like wormhole, now we are going to derive the conserved quantities for the traversable wormhole sourced by electromagnetic field:
\begin{equation}
\begin{gathered}
    \mathcal{E}=-2 \left(r \left(-2 M+\Lambda ^2 r+r\right)+Q^2\right)\\
    \times\Bigg(r^2 \sqrt{-\frac{4
   \left(r \left(-2 M+\Lambda ^2 r+r\right)+Q^2\right)}{r^2}-\left(\pm \left(2
   \sqrt{\frac{Q^2-M r}{r^2}}\right)\right)^2}\Bigg)^{-1}
   \end{gathered}
\end{equation}
\begin{equation}
\begin{gathered}
    \mathcal{L}=\left(\pm \left(2 \sqrt{\frac{Q^2-M r}{r^2}}\right)\right)^2\\
    \times\Bigg(2
   \sqrt{-\frac{4 \left(r \left(-2 M+\Lambda ^2
   r+r\right)+Q^2\right)}{r^2}-\left(\pm \left(2 \sqrt{\frac{Q^2-M
   r}{r^2}}\right)\right)^2}\Bigg)^{-1}
\end{gathered}
\end{equation}
\begin{equation}
    \Omega=\pm \frac{\pm \left(2 \sqrt{\frac{Q^2-M r}{r^2}}\right)}{2 r}
\end{equation}
In order to investigate the effective potential, normalized by the $L^2$ we want to numerically solve the equation (\ref{eq:3.11}). Results of such investigation are respectively plotted on the Figure (\ref{fig:444}). As one could notice, we assume both Damour–Solodukhin wormhole (specific case of RS wormhole with vanishing charge) and Reissner-N\"orstrom wormhole. It is obvious that for each case (DS and RS) photon sphere exists, which is validated by the effective potential peak nearby the throat and it's radial derivatives. Besides, following the same procedure that was applied for Schwarzschild wormhole, we show the null-like trajectories near the charged wormhole solution on the second plot of the Figure (\ref{fig:4}).

\section{Thin accretion disks and their optical properties}\label{sec:4}
\begin{figure*}[!htbp]
    \centering
    \includegraphics[width=\textwidth]{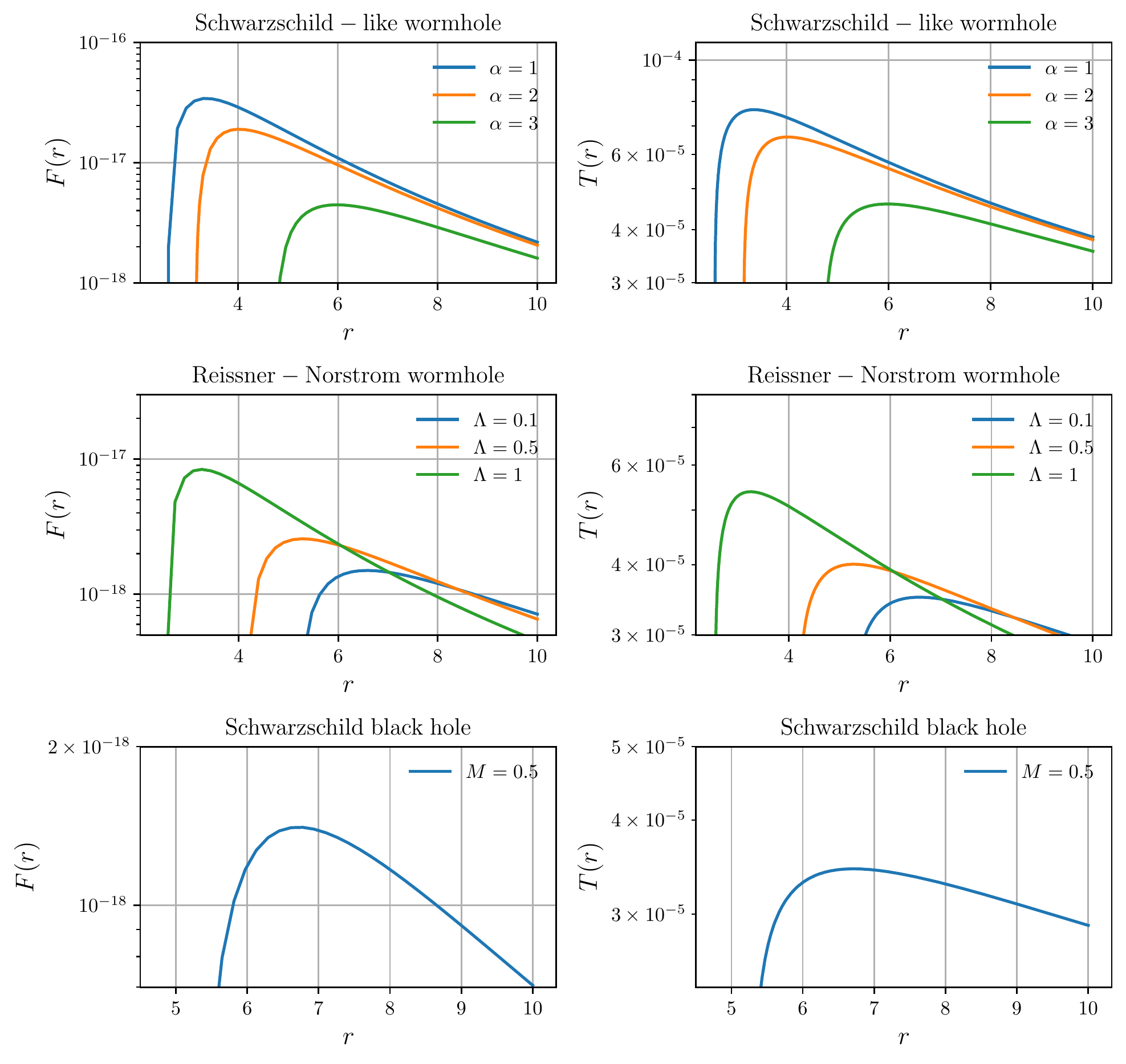}
    \caption{Radiation flux and temperature for Schwarzschild-like wormhole, Reissner-N\"ordstrom wormhole and Schwarzschild black hole for comparison with $\sigma=1$, $Q=\Lambda=0.1$, $M=0.5$, $r_0=1$ (for Schild WH) and $\dot{M}=10^{-12}M/$yr}
    \label{fig:1121}
\end{figure*}

In the following, we are going to briefly discuss the physical properties of the optically thick accretion disks. Such accretion disks occur nearby the compact objects (both of singular and non-singular nature) because of the mass transfer onto the aforementioned massive object. Because of the present matter viscosity (both shear and bulk), matter falls onto the compact object (in the case of BH solution, onto the event horizon, on the other hand, if one assumes wormhole as a compact object, matter transits through the wormhole throat into the other universe).  It is obvious that for thin accretion disks, matter content is generally distributed within the equatorial plane. As was mentioned in the work of \citep{RAHAMAN2021115548}, one could state that thin accretion disks in the ray-tracing simulations inhabited by the null-like particles with four-velocity $U^\mu$ that form a photon sphere near the wormhole throat and has an averaged surface density
\begin{equation}
    \Sigma=\int^H_{-H}\langle\rho_0\rangle dz
\end{equation}
Here, $H$ denotes the accretion disk height, such that $H/R\ll1$ and $\langle\rho_0\rangle$ denotes the averaged energy density assuming $\theta=2\pi$. Here, we used the following definition of energy-momentum tensor for the imperfect fluid (since the energy flux is present, non-diagonal components could be non-zero) \citep{Misner:1973prb}:
\begin{eqnarray}
\begin{gathered}
    T_{\mu\nu}=\rho_0 (1+\epsilon) U_\mu U_\nu + (p-\zeta\theta) h_{\mu\nu}-2\eta \Sigma_{\mu\nu}\\
    +q_\mu U_\nu + q_\nu U_\mu+t_{\mu\nu}
    \label{eq:6.4}
\end{gathered}
\end{eqnarray}
As usual, in the equation above $\eta$ and $\zeta$ are shear and bulk viscosities respectively, $\theta=\nabla_\mu U^\mu $ and $q_\mu$ is the energy flux vector, $t^{\mu\nu}$ is the stress tensor and $h_{\mu\nu}$ is the induced metric, $h$ is it's trace:
\begin{equation}
    h_{\mu\nu}= U_\mu U_\nu +g_{\mu\nu},\quad h=h^\mu_{\;\;\mu}
\end{equation}
Finally, we could also give the definition for the shear tensor:
\begin{equation}
    \Sigma^{\mu\nu}=\frac{1}{2}\bigg(\nabla_\alpha U^\mu h^{\alpha\nu}+\nabla_\alpha U^\nu h^{\alpha\mu} \bigg)-\frac{1}{3}\theta h^{\mu\nu}
\end{equation}
If the entropy for the considered fluid is conserved and shear, bulk viscosities are omitted, Equation (\ref{eq:6.4}) could be greatly simplified to (isotropic pressure vanish as well)
\begin{eqnarray}
    T_{\mu\nu} = \rho_0 h U_\mu U_\nu +q_\mu U_\nu + q_\nu U_\mu +t_{\mu\nu}
    \label{eq:6.6}
\end{eqnarray}
This is exactly the energy-momentum tensor for the perfect fluid with the present energy flow and stress. Apart from the averaged surface density, one could also compute radiation flux and disk torque for the sake of completeness \citep{RAHAMAN2021115548}
\begin{equation}
    F(r)=\langle q^z \rangle
\end{equation}
\begin{equation}
    W^r_\phi=\int^H_{-H}\langle t^r_\phi \rangle dz
\end{equation}
Here, $\langle t^r_\phi \rangle$ denotes the averaged off-diagonal component of the stress tensor. Finally, plugging everything up (with the use of energy-momentum tensor and baryon conservation equations), one could obtain the exact expression of the radiation flux, the most fundamental quantity of thin accretion disks \citep{1974ApJ...191..507T,RAHAMAN2021115548}
\begin{equation}
    F(r)=-\frac{\dot{M}}{4\pi\sqrt{-g}}\frac{\partial_r\Omega}{(\mathcal{E}-\Omega\mathcal{L})^2}\int^r_{r_{\mathrm{ms}}}(\mathcal{E}-\Omega\mathcal{L})\partial_r\mathcal{L}dr
    \label{eq:40}
\end{equation}
\begin{equation}
    \dot{M}=-2\pi \sqrt{-g}\Sigma U^r=C
\end{equation}
where $\dot{M}$ is the mass accretion rate, which is for our case is constant and $r_{\mathrm{ms}}$ is the radius of the marginally stable orbit, which for Schwarzschild and RS wormhole equals to
\begin{equation}
    r_{\mathrm{ms}}=\alpha+\frac{\alpha}{\sqrt{3}}
    \label{eq:43}
\end{equation}
\begin{equation}
    r_{\mathrm{ms}}=\frac{4 M}{\Lambda ^2+1}-\frac{2 Q^2}{\Lambda ^2+1}
        \label{eq:4444}
\end{equation}
Several quantities could be additionally computed, such as thin accretion disk temperature \citep{TORRES2002377}:
\begin{equation}
    F(r)=\sigma T^4
    \label{eq:44}
\end{equation}
where $\sigma$ is the usual Boltzmann constant, here we assume that it does equals to unity for the sake of simplicity. Therefore, we numerically solve Equations (\ref{eq:40}) and (\ref{eq:44}) with the use of numerical integrator for Schwarzschild, Reissner-N\"ordstrom wormholes and for Schwarzschild black hole in order to compare the results. To produce the plots, we as usual assume that the Schwarzschild-like wormhole radius is unitary, RS $M=0.5$ and charge $Q=0.1$. Moreover, Damour–Solodukhin parameter $\Lambda=0.1$. For Schwarzschild black hole we use the same mass $M=0.5$ as was used for both RS and Schwarzschild-like wormholes. Results for radiation flux and black-body temperature are given in the arbitrary units on the Figure (\ref{fig:1121}). Now we are going to briefly discuss the results for each spacetime kind:

\textit{Schwarzschild wormhole}: this is the first wormhole model of our consideration. For that solution, one could easily notice that radiation flux maximum shifts to the $r\to\infty$ if we assume relatively big values of $\alpha$ parameter, which coincides well with the effective potential from the Figure (\ref{fig:3}). Same behavior could be observed for the black body temperature.

\textit{Reissner-N\"ordstrom wormhole}: second exotic wormhole spacetime, and the last that we are going to examine. For that case, it is obvious that $\max F(r)\to0$ as $\Lambda\to\infty$. Therefore (similar to the Schwarzschild wormhole), radiation flux of the thin accretion disk match the $V_{\mathrm{eff}}$. However, it could be remarked that both radiation flux and temperature has smaller values in relation to the first wormhole model.

\textit{Schwarzschild black hole}: simplest black hole model admitting spherical symmetry added for comparison. For the SS BH solution, we could note that generally radiation flux $F(r)$ and black-body temperature of the accretion disk $T(r)$ are smaller than the ones, that was derived for wormhole models of our consideration (however, results match with the RS wormhole for vanishing Damour–Solodukhin parameter and electromagnetic charge, as expected).

Since we already investigated all of the necessary thin accretion disk properties numerically and graphically, now we are going to proceed further and obtain the images of such accretion disks nearby two wormhole models of our consideration with the use of ray-tracing techniques, for example explained in details in the pioneering works of \citep{Paul_2020,Shaikh_2019}. In order to obtain those images, we will properly modify the open-source ray tracing code \texttt{GYOTO} (this code was described in the papers \citep{Vincent_2011,Vincent_2016}, documentation is provided on \href{http://gyoto.obspm.fr}{http://gyoto.obspm.fr}). Images were obtained from the point of view of the asymptotical observed (we assume that observer plane lies within $\theta=\pi/2$ (equatorial region) and that $r_{\mathrm{obs}}=500r_0$). Additionally, as it could be seen from the value of $r_{\mathrm{obs}}$, we assume the only one case, when the asymptotical observer located at the same universe as the accretion disk (such assumption has been made for the sake of \texttt{GYOTO} output validity).
\begin{figure*}[!htbp]
    \centering
    \includegraphics[width=\textwidth]{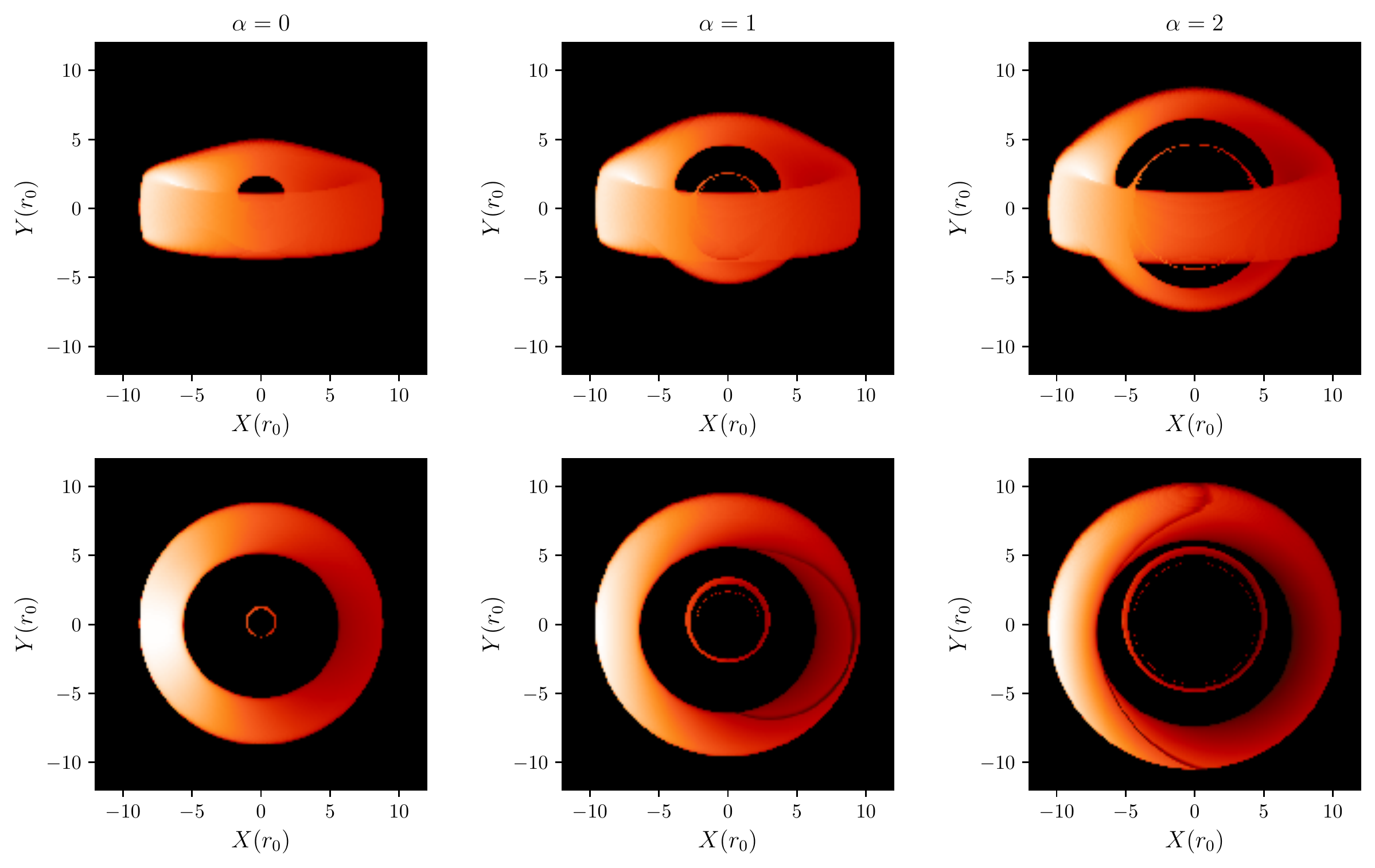}
    \caption{Ray-tracing simulations of an accreting optically thin torus around the Schwarzschild-like wormhole with varying $\alpha$}
    \label{fig:174}
\end{figure*}

\subsection{Model I}\label{sec:4.1}
We plot the ray-tracing images of the accreting tori (which is optically thin) for Schwarzschild-like wormholes with the varying (and positive) values of $\alpha$ on the Figure (\ref{fig:174}). In order to produce aforementioned images, we used inclination angles $i=105^{\circ}$ and $i=20^{\circ}$. Second angle ($i=20^{\circ}$) has been chosen in order to produce the images, for which black-hole and disk spin vectors are directed "into the page" (for more details on the subject, see papers \citep{2018ApJ...855..128W,EventHorizonTelescope:2019pgp,Vincent2021}). Now, we are going to investigate each image of Schwarzschild wormhole.

\textit{First case}: $\alpha=0$. This case refers to the vanishing redshift function, and therefore Zero Tidal Forces (ZTF) wormhole, since for our spherically symmetric line element, redshift function acts as a gravitational potential. As we already depicted on the Figure (\ref{fig:3}), radius of the photon sphere (which is defined as a last stable orbit, formed by a null-like particles, that could oftenly act as the inner radius of an accretion disks nearby the compact objects) decrease as $\alpha\to0$, that could also be noticed on the first and fourth plots of the Figure (\ref{fig:174}). From the ray-tracing simulations, we could conclude that photon sphere does exist for the case with vanishing tidal forces, however it could be completely indistinguishable if we assume EHT resolution (which equals approximately to $\approx20\mu$as, reported by \citep{2019ApJ...875L...4E}). On the aforementioned plot, one could obviously identify the light ring (i.e. photon sphere) and thick annular area (as it could be seen from the polar region). Thick annular region corresponds to the optically thin accreting tori. 

\textit{Second case}: $\alpha=1$. As the second case, we consider redshift function $\Phi=-1/r$, that corresponds to the value of $\alpha=1$. Here, redshift function does not vanish, but because of the asymptotical flatness condition $\Phi\to0$ as $r\to\infty$, as required for the viability of the theory. For the second case, as well as it was for the previous one, photon sphere could be recognised by the asymptotically far located ZAMO (Zero Angular Momentum Observer) as well as the thick annular area (polar projection of the thin accretion disk with toroidal nature). Besides, for unblurred images one could notice that secondary, faint ring appears near the photon sphere, which could be referred to as a product of lensing. However, unfortunately secondary ring could not be observed by the current VLBI interferomenters, but could be in the future (for example, using extended EHT2025 configuration).

\textit{Third case}: $\alpha=2$. The last value of $\alpha$ that we consider in the present study is $\alpha=2$, for which redshift function has exact form $\Phi=-2/r$. In the previous subsections, when we investigated the Schwarzshild wormhole orbital mechanics with the help of effective potential $V_{\mathrm{eff}}$ (for more detailed graphical representation, see Figure (\ref{fig:3})), this, third case had $r_{\mathrm{ph}}=2r_0\mathcal{L}^2$, which is validated by the ray tracing simulations on the third and sixth plots of the Figure (\ref{fig:174}). For $\alpha=2$ and bigger values of $\alpha$ secondary light ring could be recognised more easily, since it's radius grows with $\alpha\to\infty$. Moreover, inner radius of accreting tori grows with increasing $\alpha$.

Considering that we already discussed all three cases for Schwarzschild wormhole, we could go ahead to the second wormhole model of our consideration, namely Reissner-N\"ordstrom wormhole (in the following subsection, we will investigate both RS wormholes with positive charge and special case, namely DS wormhole with vanishing electromagnetic charge). 
\begin{figure*}[!htbp]
    \centering
    \includegraphics[width=\textwidth]{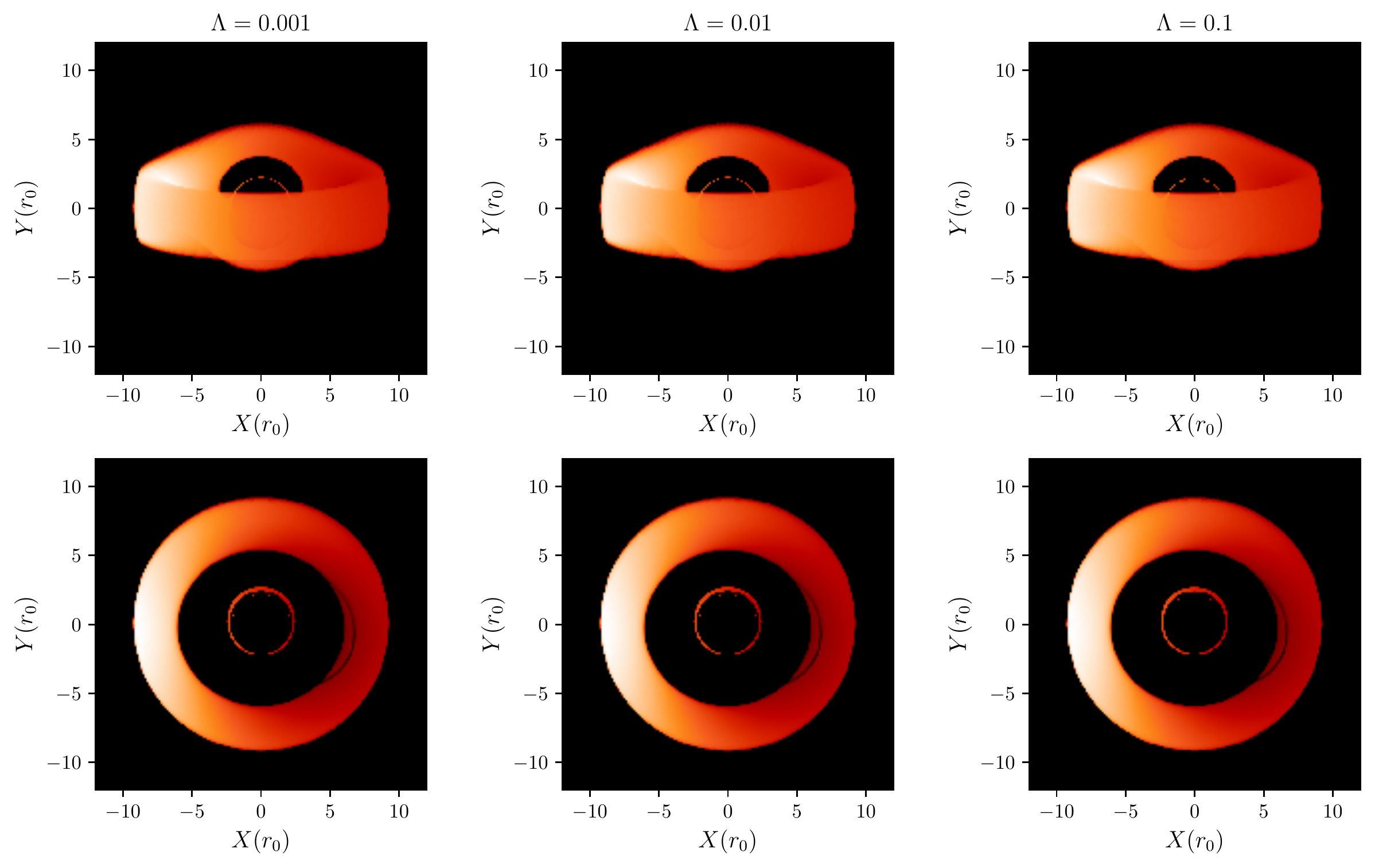}
    \caption{Ray-tracing simulations of an accreting optically thin torus around the charged Reissner-N\"ordstrom wormhole with $Q=0.1$, $M=0.5$ and varying $\Lambda$}
    \label{fig:175}
\end{figure*}

\subsection{Model II}
The results of the ray tracing near Reissner-N\"orstrom wormhole are respectively plotted on the Figure (\ref{fig:175}) with the different values of $\Lambda$ parameter and inclination (we have chosen two cases with $i=105^{\circ}$ and $i=20^{\circ}$). This choice of inclination was elaborated in the previous subsection. Consequently, following the same procedure, that was applied for the Schwarzscild wormholes, we now are going to discuss the ray-tracing images, produced for $\Lambda=10^{-3},10^{-2},10^{-1}$ respectively:

\textit{First case}: $\Lambda=0.001$. For that option, Reissner-N\"orstrom wormhole tidal forces are the smallest among other choices of $\Lambda$ in our paper. Moreover, gravitational potential $\Phi$ vanish at the asymptotical infinity and satisfy both conditions of asymptotical flatness and horizonless. Consequently, as it was prescribed by the effective potential and it's derivatives from the Figure (\ref{fig:444}), radius of the photon sphere $r_{\mathrm{ph}}$ is the biggest and shrink for $\Lambda\to\infty$, for vanishing $\Lambda$, $r_{\mathrm{ph}}$ coincides with the RS black hole case. Moreover, for the relatively small and positive values of $\Lambda$ parameter secondary lensed light ring appears nearby the photon sphere.

\textit{Second case}: $\Lambda=0.01$. This is the second case of our consideration for RS-like wormholes. Now, our wormhole solution differ from the RS black hole more, which could be confirmed by the smaller radius of the photon sphere. However, secondary lensed ring still could be observed (for EHT resolution of 20 $\mu$as, secondary ring will be absorbed into the main, photon ring because of the big difference in flux).

\textit{Third case}: $\Lambda=0.1$. This is namely the last case of our consideration that has the smallest radius of the photon sphere within the chosen values of $\Lambda$ and therefore tidal forces of such wormhole solution are the biggest, which could be the answer for the photon ring shift up to the wormhole throat. Remarkably, for $\Lambda\gg0$ one could observe that secondary lensed ring disappear.

Every feature of the ray-tracing images mentioned above holds for the special case with vanishing electromagnetic charge (Damour-Solodukhin wormhole) except that radius of the photon ring for each case is slightly smaller.

Therefore, we discussed each case for both Schwarzschild-like and Reissner-N\"orstrom wormholes and we could move to the next section, where we will investigate the geometrically thick accretion disks and process obtained images in order to achieve the EHT VLBI resolution (20 $\mu$as).
\begin{figure*}[!htbp]
    \centering
    \includegraphics[width=\textwidth]{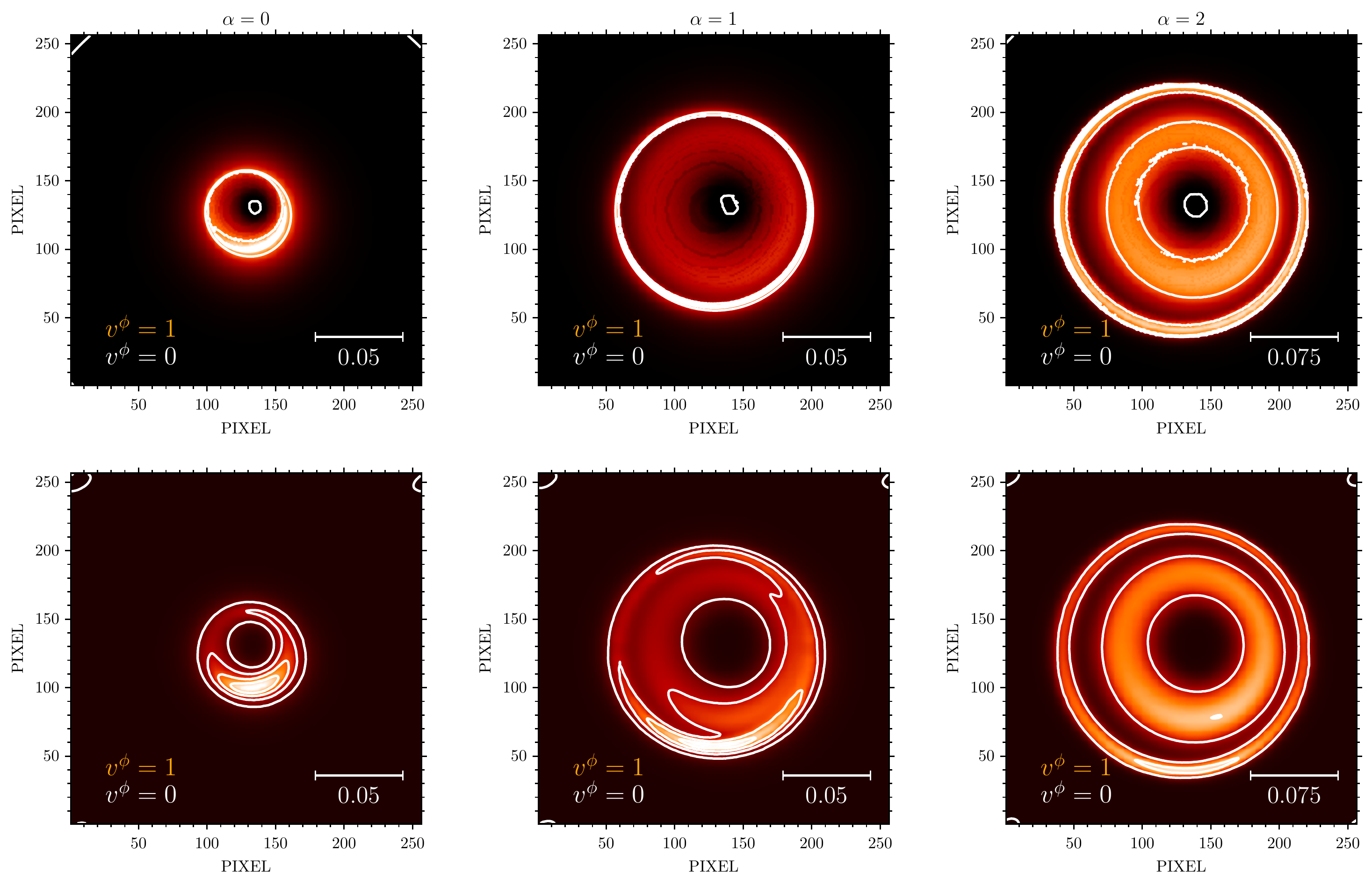}
    \caption{Ray-tracing images of the Schwarzschild-like wormhole surrounded by the thick accretion disk polar regions with varying $\alpha$. On the Figure, we depict both rotating and radially plunging flows (with the use of contours)}
    \label{fig:12}
\end{figure*}

\section{Thick disk and EHT imaging}\label{sec:5}
As we already mentioned, in the present section we are going to study the geometrically thick (optically thin) accretion disks around the exotic wormhole models that we consider. In the \texttt{GYOTO} package, thick accretion disks are generally parameterized by two fundamental quantities, namely disk inner radius (in our paper, we set $r_{\mathrm{in}}=r_0$ in order to reproduce physically viable case) and disk opening angle $\theta_{\mathrm{op}}$, that defines the angle between the equatorial plane $\theta=\pi/2$ and accretion disk surface (it is worth to notice that there is no outer disk radius implemented, since it is depends on the chosen spectrograph, maximal radial coordinate value $r_{\mathrm{max}}$). As it was already emphasised in the paper \citep{Vincent2021}, to produce optically thin ray-tracing images of an accretion disk, one must assume that $\theta_{\mathrm{op}}=30^{\circ}$. In the open source package \texttt{GYOTO}, that we use, one could also assume the magnetisation of the accretion disk with the help of magnetisation parameter $\sigma=\mathcal{B}^2/4\pi/(m_pc^2n_e)$ (here we define $\mathcal{B}$ as a magnetic field amplitude, $m_p$ - proton mass and $n_e$ - electron number density). We use the value of magnetisation parameter $\sigma=0.01$, which is set by default in the \texttt{GYOTO} package. As well, we fix electron number density and accretion disk temperature at the inner radius to $n_e(r_{\mathrm{in}})=2.4\times10^5$cm$^{-3}$ and $T(r_{\mathrm{in}})=8\times10^{10}$K respectively. Finally, it is convenient to remark that we will use the following expression for the accretion flow four-velocity (measured by the ZAMO):
\begin{equation}
    \mathbf{V}=V^r\frac{\partial_r}{\sqrt{g_{rr}}}+V^r\frac{\partial_\phi}{\sqrt{g_{\phi\phi}}}
\end{equation}
So that both timelike and $V^\phi$ four-velocity components vanish. In favor of circularly rotating accreting flow, one must assume purely azimuthal velocity, and therefore 
\begin{equation}
    v^\phi=V^\phi/V=1
\end{equation}
where
\begin{equation}
    V^2=(V^r)^2+(V^\phi)^2
\end{equation}
On the other hand, one could also produce a ray-tracing image of the radially plunging flow if $U^\phi=0$.
\begin{figure*}[!htbp]
    \centering
    \includegraphics[width=\textwidth]{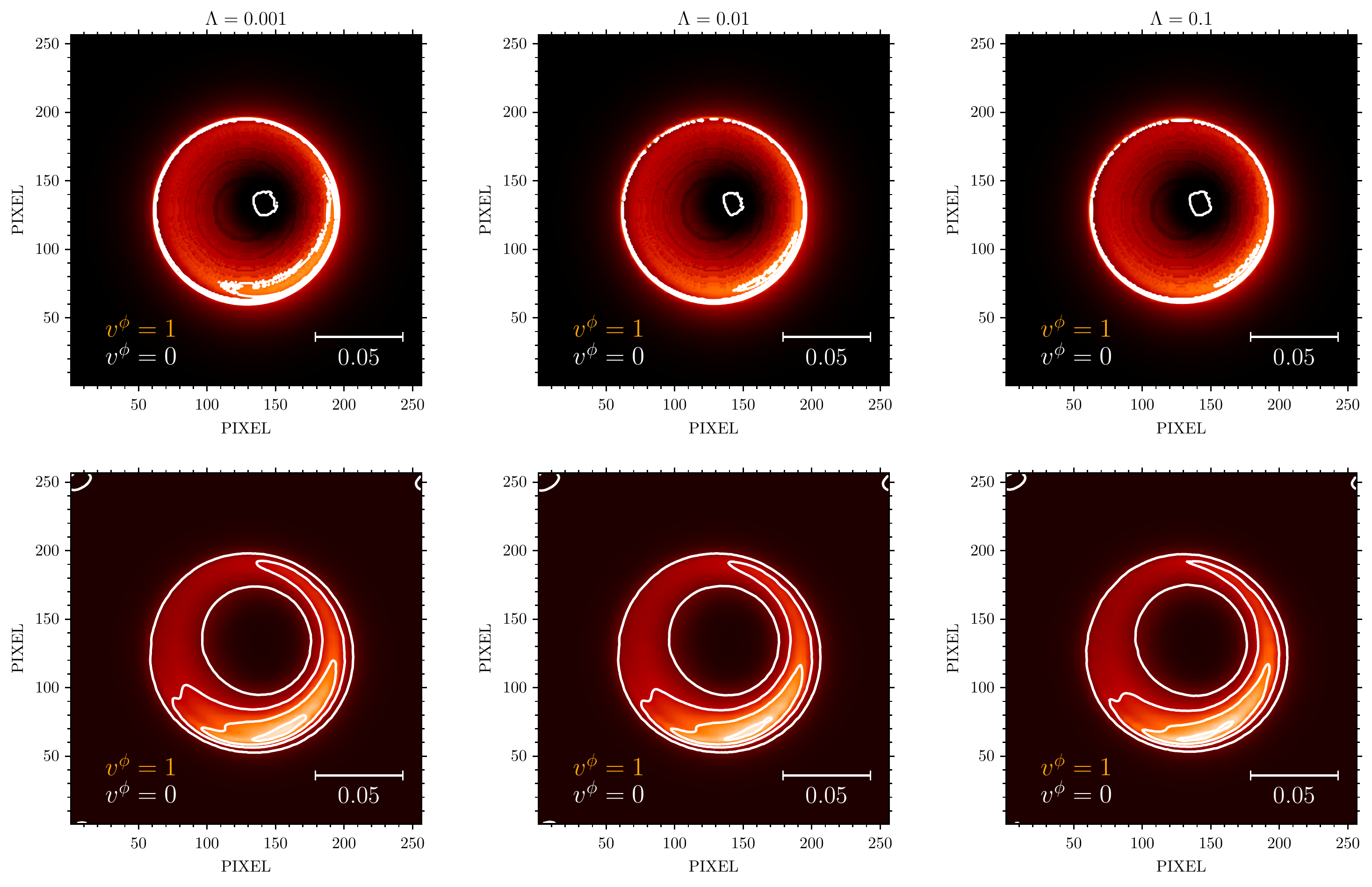}
    \caption{Ray-tracing images of the Reissner-N\"ordstrom wormhole surrounded by the thick accretion disk polar regions with $Q=0.1$ and varying DS parameter $\Lambda$, rotating accretion flow ($v^\phi=1$). As well, using contours we depict plunging accretion flows ($v^\phi=0$)}
    \label{fig:13}
\end{figure*}
\subsection{Model I}
As usual, we will start our investigation from the first wormhole solution, namely Schwarzschild-like wormhole with varying gravitational potential $\Phi$. In relation to the thin accreting torus, here we only consider one value of inclination, $i=20^{\circ}$. Such value could provide the complete information about the wormhole configuration (radius of the photon sphere and secondary ring). Additionally, our image coincides with the EHT VLBI network point of view (besides, all of the images, that will be listed below are obtained in the EHT band $\nu=230$GHz). Moreover, in order to reproduce the images, blurred up to the EHT resolution, we use package \texttt{ehtim} (for the detailed description of the package and it's capabilities, see paper \citep{2016ApJ...829...11C,2018MNRAS.478.5209C}, open source package could be downloaded via the link \href{http://github.com/achael/eht-imaging}{http://github.com/achael/eht-imaging}), which works on the basis of regularized maximum likelihood. Therefore, we are going to provide the key points on the each case ($\alpha=0,1,2$) below:

\textit{First case}: $\alpha=0$. As it was beforehand stated, vanishing $\alpha$ corresponds to the Zero Tidal Forces Schwarzschild wormhole. Only for the ZTF case, one could notice the well-defined boost towards the observer, that arise because of the special relativistic effects, Doppler shift (accretion flow with the circular rotation that coming towards the observer causes the anisotropic distribution of the flux density within the thick accretion disk, which agrees greatly with the EHT observations). Besides, such feature as aforementioned boost is present for the non-vanishing tidal forces, but not as strongly marked as for ZTF case.

\textit{First case}: $\alpha=1$. For non-ZTF cases, photon ring could be easily differentiated from the thick accretion disk even at the EHT resolution. As expected, photon ring has bigger angular diameter but secondary ring could not be resolved.

\textit{First case}: $\alpha=2$. This is the last value of $\alpha$ that we use to probe thick accretion disk. For that case, we observe that secondary ring appears. Besides, as one could see, now photon ring completely decouples from the thick accretion disk, but secondary ring is absorbed by the photon ring on the blurred image with low resolution. We could approximately constrain the values of $\alpha$ to be relatively small and positive, since it was reported by the EHT collaboration that asymmetric ring-like structure were observed at the radius of 40$\mu$as \citep{EventHorizonTelescope:2019pgp}. GRMHD simulations of various wormhole models show that this ring-like feature could be the well-known photon ring and it's asymmetry is caused by the black hole spin \citep{EventHorizonTelescope:2019pgp}.

\subsection{Model II}
The second wormhole model is as usual spherically symmetric wormhole preserving electromagnetic charge. Ray-tracing simulation snapshots for the aforementioned wormhole model are consequently placed on the Figure (\ref{fig:13}). It could be seen that radius of the photon ring does decrease with the growing values of $\Lambda$ (which as well signifies the difference between our wormhole model and RS charged black hole). As it was already mentioned, EHT has observed a ring-like feature with angular diameter of approximately 40 $\mu$as, which could lead for our model to the relatively big and positive values of $\Lambda$ in order for the model to match with the present VLBI observations.
\subsection{Radially plunging flow}
Here, in relation to the previous subsections, we are going to assume the radially plunging flow that is defined by $v^\phi = V^\phi/V=0$ so that $V^\phi$ vanish.

We place the contours of ray tracing simulation snapshots with vanishing azimuthal velocity on the Figures (\ref{fig:12}) and (\ref{fig:13}) for both Schwarzschild and Reissner-N\"ordstrom wormholes for the sake of comparison between $v^\phi=1$ and $v^\phi=0$ cases. As we see, practically pictures of radially plunging accretion flows does not significantly deviate from the ones with rotating matter content ($V^r=V^\phi\neq0$).
\section{Imaging accretion jet}
Astrophysical compact objects are often associated with collimated, relativistic jets of matter expelling from the objects close to the speed of light. These jets are associated with strong magnetic fields whose underlying General Relativistic Magnetohydrodynamic (GRMHD) physics gives insights to the dynamics of the accretion disk, associated compact object, stellar progenitor, and the object environment which are relevant for studies associated with Super Massive Black Holes (SMBHs) and Active Galactic Nuclei (AGN) to understand galaxy and stellar evolution since they have been shown affect star formation rates, aiding and constraining star formation by supplying energy into surrounding Interstellar Medium (ISM) \citep{mizuno2022grmhd} \citep{ressler2021magnetically}. The formation and destruction of these jets itself is an active area of study to understand astrophysical turbulence and objects such as Neutron Stars and X-Ray Binaries (XRBs), but those are not under the purview of this current study and one may refer to \cite{ressler2021magnetically} or \cite{kylafis2012formation} for further information. 

Ray-traced images of Black Holes, Wormholes and other compact objects, similar to the ones we have obtained thus far, show features across a large range of length scales. These range from extremely fine structures observed in accretion tori and the photon rings, to large scale coherent structures observed in the relativistic jets associated with such objects. In this section we aim to focus on imaging and observing the latter by analysing ray-traced image of jet-thick accretion disk system around exotic wormhole models. This is similar in part to \cite{kramer2021ray} who use radioactive transfer calculations on 3D GRMHD simulations to understand jet dynamics solely. We extend their methods to study jets associated with thick-disk exotic wormholes. These jets are generally found to coexist with optically thick accretion disks, commonly observed in SMBHs found at galaxy centres, some white dwarfs, and Neutron Stars; and have been hypothesised to exist with wormholes too \citep{kirillov2020wormhole}. 

Here and further throughout this section we are going to construct the model of thick accretion disk with the present jet in order to obtain the ray-tracing image, similar to the M87. We are going to introduce the physical properties of the simple analytical jet model, discussed in details within the pioneering works of \citep{Monika2013,Davelaar2018}. In those aforementioned papers, GRMHD simulations of accretion relativistic jets were set up and it was unveiled that generally inner parts of the jet (the so-called spine) does not contain emitting matter and therefore emission comes only from the outer jet regions (jet sheath).

Within the \texttt{GYOTO} disk-jet model, there is a large variety of free parameters, that define jet accretion properties present at the moment. For example, the emitting region of the jet is defined by two angles $\theta_1$ and $\theta_2$. Since this emission layer is assumed to be thin, $\theta_2$ is only slightly bigger that $\theta_1$. Moreover, there is present the so-called jet base height $z_b$. Following the work of \citep{Vincent2019}, we set that $\theta_1=\theta_2-3.5=20^{\circ}$ and that jet base height does equals to the wormhole throat radius $z_b=r_0$. Moreover, here and further we assume that the bulk Lorentz factor measured by ZAMO is $\Gamma=1.15$ (that is set by default in the simulation suite). In order to properly ray-trace our exotic wormhole spacetimes, we need to define the exact form of jet particles four velocity \citep{Vincent2019}:
\begin{equation}
    \mathbf{U}_{\mathrm{jet}}=\Gamma( \mathbf{U}_{\mathrm{ZAMO}}+ \mathbf{V})
\end{equation}
Here $\mathbf{U}_{\mathrm{ZAMO}}$ is the four-velocity of the observer with vanishing angular momentum that is located at the asymptotically flat region and $\mathbf{V}$ is the velocity of relativistic jet measured by ZAMO (with vanishing time-time and $\phi\phi$ components). This velocity has only two non-vanishing components (namely radial and $\theta\theta$) because of the relativistic jet axial symmetry and stationary nature. Besides, it will be more appropriate to write down the jet velocity and it's non-vanishing components in the terms of sheath angles $\theta_1$ and $\theta_2$:
\begin{equation}
    \mathbf{V}=V(\sin \theta_k \mathbf{e}_r +\cos \theta_k e_z)
\end{equation}
Where $\mathbf{e}_r$ and $\mathbf{e}_z$ are respectively radial and azimuthal unit vectors, $V=\sqrt{\Gamma-1}/\Gamma$ is the jet velocity given in the units of light speed. In the equation above, $\theta_k$ could be both equal to inner, outer opening angles. Since we already mentioned all of the necessary theoretical foundations of accretion jet simulations, we could proceed to the investigation of such structures, that appear around exotic traversable wormholes.
\subsection{Model I}
\begin{figure*}[!htbp]
    \centering
    \includegraphics[width=\textwidth]{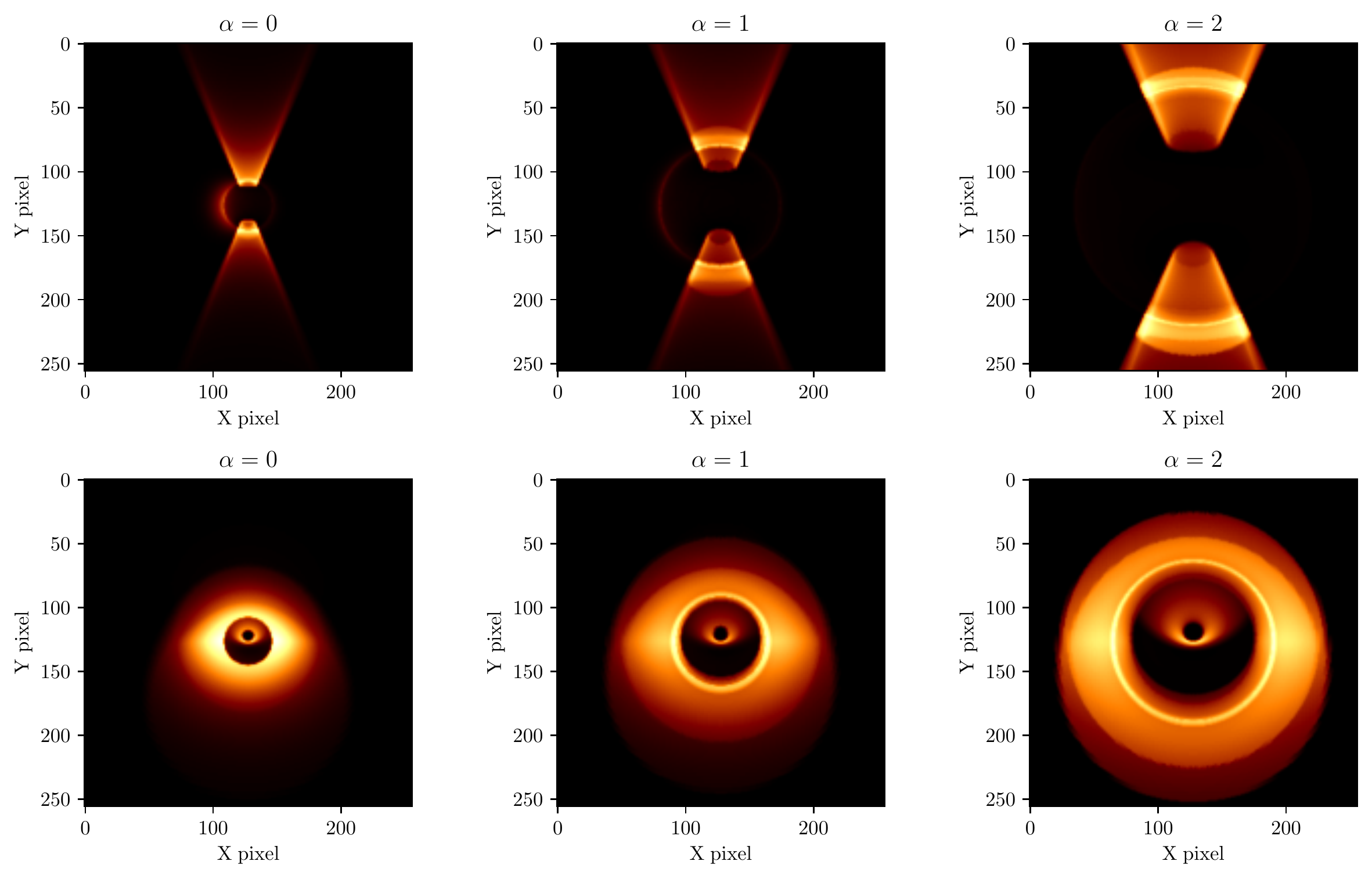}
    \caption{Ray-tracing simulations of an accreting optically thick disk with relativistic jet around the tidal Schwarzschild-like wormhole with $r_0=1$ and varying $\alpha$}
    \label{fig:1112}
\end{figure*}

The first traversable wormhole model of our consideration that we are going to discuss is as usual Schwarzschild-like wormhole with non-vanishing gravitational potential. Numerical results are routinely represented on the Figure (\ref{fig:1112}) with the varying values of free parameter $\alpha$ (i.e. varying tidal forces strength). Firstly, one could discuss the vanishing $\alpha$ case:

\textit{ZTF wormhole}: as it was already observed, for Zero Tidal Forces (ZTF) traversable wormhole solution with constant shape function, photon sphere has the smallest radii (and therefore, accretion disk $r_{\mathrm{in}}\approx r_{\mathrm{ph}}$ is relatively small). These accretion disk features were recognised in the ray tracing image for disk-jet system as well. However, here one could also see that the actual size of relativistic jet is as well smaller in relation to the other cases with $\alpha>0$, which is expected.

\textit{$\alpha=1,2$ case}: on the other hand, for the non-vanishing gravitational potential $\Phi(r)=-1/r$ (or $\Phi(r)=-2/r$) we have that accretion disk become more faint in relation to the relativistic jet brightness. As was previously noted, angular diameter of the jet grows as well as the photon sphere radius. Moreover, for inclination angle $i=70^{\circ}$ photon sphere could be easily noticed even with the EHT resolution.

In all of the cases, as it could be easily remarked, brightness of the relativistic jet dominates over the accretion disk luminosity. However, it could be fixed with the assumption of bigger inner temperature of the disk $T_{\mathrm{in}}$. There also exist the brightest area of the relativistic jet, located nearby it's base, that could appear because of the photon bending and it's radius is approximately equals to the radius of the photon sphere.

\subsection{Model II}
\begin{figure*}[!htbp]
    \centering
    \includegraphics[width=\textwidth]{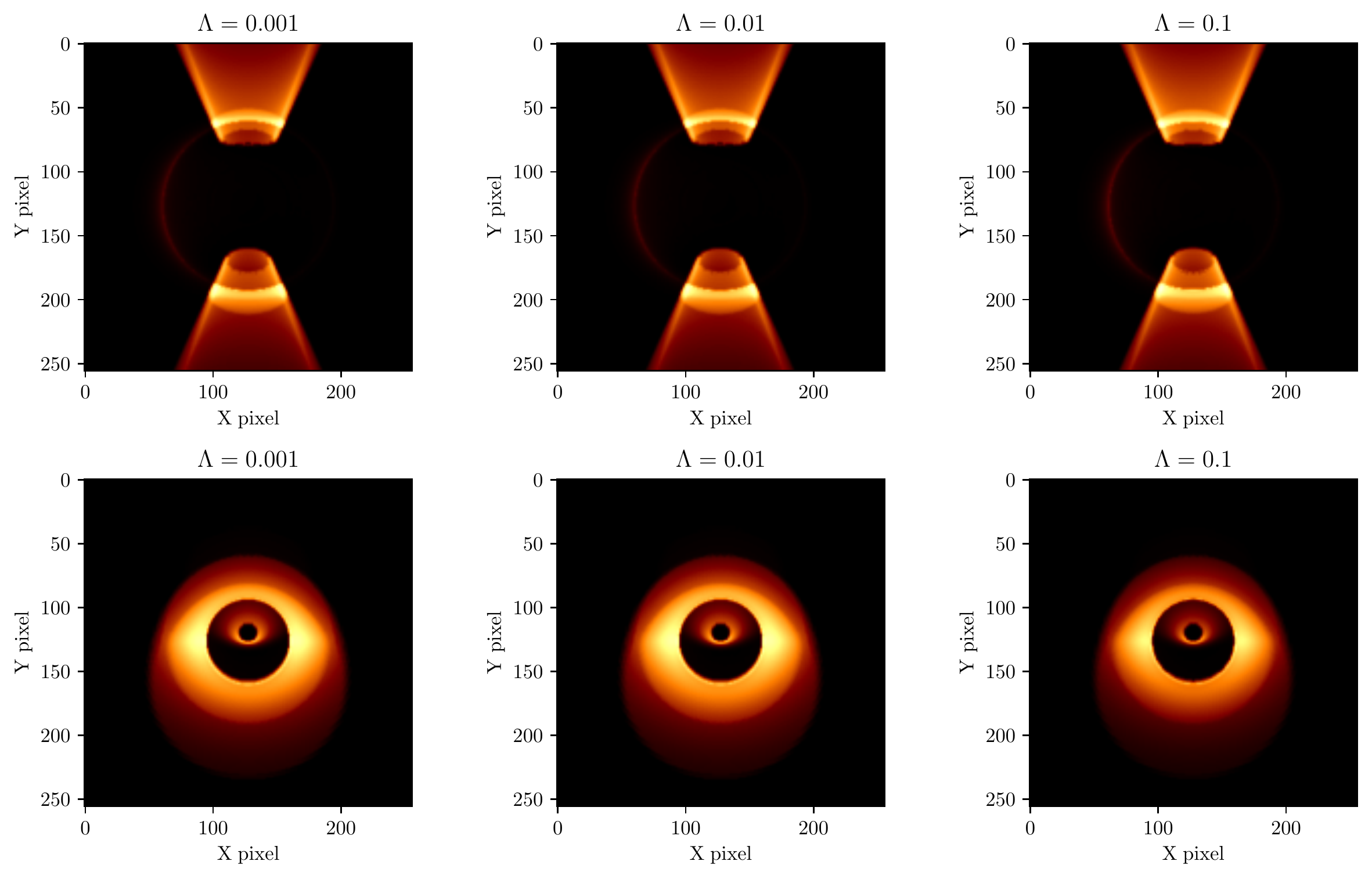}
    \caption{Ray-tracing simulations of an accreting optically thick disk with relativistic jet around the tidal Reissner-N\" ordstrom wormhole with $M=0.5$ and $Q=0.1$ and varying $\Lambda$}
    \label{fig:1113}
\end{figure*}

Commonly, our second and last wormhole model is charged RS WH. Ray-tracing images of relativistic jet and accretion disks for RS WHs are placed on the Figue (\ref{fig:1113}). As one  could obviously notice, not only photon sphere, relativistic jet gets slightly smaller with the $\Lambda\to\infty$, but as well the disk and relativistic jet become more faint with the growing values of DS parameter.
\section{Microlensing of the star bulb}\label{sec:6}
The last ray-tracing scenario that we are going to study is namely the microlensing of the star bulb (radiative fluid sphere), caused by the wormhole curvature. In order to produce the ray-tracing images, we used the \texttt{GYOTO} fixed star scenario and placed the star bulb at $r_{\mathrm{st}}=10r_0$ assuming that the star radius is $R=4$ (for Schwarzschild-like wormhole) and $R=6$ (for Reissner-N\"odstrom wormhole). With the help of microlensing, one could see how the wormhole shadow changes with the change of $\alpha$, $\Lambda$ values. As well, we varied the value of $\phi$ for a fixed star from $\phi=\pi$ up to the $\phi=\pi/2$.
\begin{figure}[!htbp]
    \centering
    \includegraphics[width=\columnwidth]{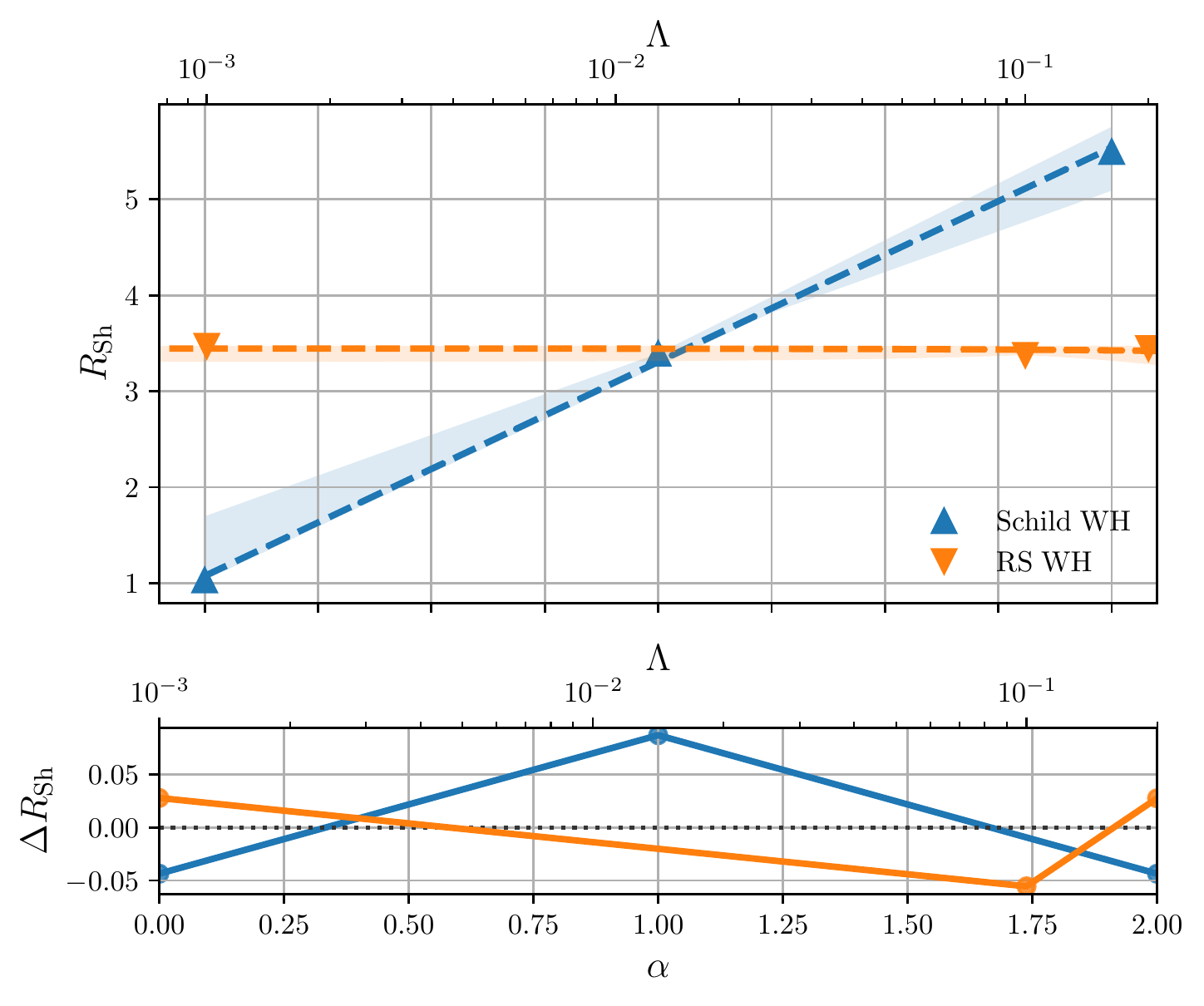}
    \caption{Radius of the Schwarzschild-like (blue triangles) and Reissner-N\"ordstrom (orange triangles) wormhole shadow with varying $\alpha$ (lower axis) and $\Lambda$ (upper axis) with the present linear regression fits}
    \label{fig:prelast}
\end{figure}
\subsection{Model I}
Main feature that we are going to determine for the Schwarzschild wormhole is the radius of the wormhole shadow. Approximate values of wormhole shadow radius $R_{\mathrm{sh}}$ for $\alpha=0,1,2$ is plotted on the Figure (\ref{fig:prelast}). It could be seen that radius of the wormhole shadow grows linearly with $\alpha$, which is also validated by the linear regression that fits perfectly for the obtained dataset (we as well plot the 95\% confidence intervals on the aforementioned figure):
\begin{equation}
     R_{\mathrm{Sh}}=2.232 \alpha + 1.0745
\end{equation}

\subsection{Model II}
Here we are going to briefly discuss the Reissner-N\"ordstrom wormhole shadows. It could be generally said that radius of the RS wormhole shadow does shrink with the growing values of $\Lambda$, which coincide with the decreasing radius of the photon ring, that could be seen on the Figure (\ref{fig:13}). Linear regression fit for our dataset is given below as follows:
\begin{equation}
    R_{\mathrm{Sh}}=-0.1181 \Lambda + 3.445
\end{equation}

\section{Concluding remarks}\label{sec:7}
In this article we comprehensively investigate the accretion flows around two exotic wormholes with the present tidal forces, namely Schwarschild-like wormhole with constant shape function and Reissner-N\"ordstrom wormhole with the present electromagnetic field respecting $U(1)$ local gauge symmetry (for the sake of completeness, we investigate both RS wormholes and special case with vanishing electromagnetic charge, namely Damour-Solodukhin wormhole). In order to examine the accretion flows around aforementioned spherically symmetric wormhole solutions, we used effective potential and ray-tracing methods, applied for both geometrically thin and thick accretion disks.

Firstly, as we already mentioned we used the normalized by $\mathcal{L}^2$ effective potential to derive the radius of photon sphere and observe it's dynamics with the change of free parameters (additional degrees of freedom) $\alpha$ and $\Lambda$. It is worth to notice that for the Schwarzschild wormhole we considered both ZTF (Zero Tidal Forces) case and the case with the present tidal forces. The detailed discussion on the test particle orbital mechanics and the derivation of $V_{\mathrm{eff}}$ could be respectively found in the Section (\ref{sec:3}). On the other hand, for the graphical representation of the $V_{\mathrm{eff}}$ numerical solutions, refer to the Figures (\ref{fig:3}) and (\ref{fig:4}). In this subsection we as well plot the particle orbits around both WH models of our consideration, obtained with the help of lensing, direct and photon ray tracing on the Figure (\ref{fig:444}).

Secondly, we introduce the thin accretion disks and their optical properties in the next section. As well, we derive the radiation flux and temperature for Schwarzschild-like and Reinssner-N\"ordstrom wormholes, Schwarzschild black hole surrounded by the optically thin accretion disk. It was found that generally, both radiation flux and temperature for the simpliest black hole model that impose spherical symmetry and has the same mass as WH solutions is smaller. Also, behavior of the radiation flux and temperature for both wormhole models coincide well with the effective potential. Numerical solutions, obtained with the help of \texttt{Mathematica} numerical integrator \texttt{NIntegrate} for $F(r)$ and $T(r)$ are placed on the Figure (\ref{fig:1121}).

Finally, starting from the subsection (\ref{sec:4.1}) we used the numerous ray-tracing methods, that were applied to our exotic wormhole models. Ray-tracing solutions for our analytical calculations give us images which can be compared against observations from EHT and related surveys. These synthetic observations allow us to compare features between backgrounds of naked singularities, black holes, and wormholes, based on images of accretion disk with the associated photon sphere. We calculate these images based on numerical construction techniques developed in the open source massively parallel ray-tracing code \texttt{GYOTO}. These numerically constructed images allow us to compare our models qualitatively based on features found in the corresponding photon spheres and accretion tori, giving us further insights into accretion behaviour around these highly compact objects. In order to perform the ray tracing for our studies, we properly modified \texttt{GYOTO}. All of the ray-tracing simulation snapshots were obtained on the Sharanga HPC cluster nodes. Firstly, we used the optically thin accreting tori with the present electromagnetic field and magnetisation parameter $\sigma=0.01$ as a ray-tracing scenario. It was found out that for  accreting tori if we assume Schwarzschild-like wormhole as a background massive and compact object, photon sphere exists even for the case with vanishing gravitational potential (however, with the EHT resolution of $20\mu$as, it could possibly be not recognised). On the other hand, predicted by the effective potential, radius of the photon sphere shits up to the radial infinity if $\alpha\to\infty$ (see ray tracing images with varying $\alpha$ on the Figure (\ref{fig:174})). In the interest of accreting tori ray-tracing images around RS wormholes, we fix electromagnetic charge to $Q=0.1<M=0.5$ (which is the physically plausible choice). Therefore, we left with only one free parameter, namely Damour-Solodukhin parameter $\Lambda$, that signifies the difference between charged wormhole and Schwarzschild black hole. We vary DS parameter $\Lambda$ on the Figure (\ref{fig:175}). After the careful numerical examination of the obtained ray-tracing images, we could conclude that radius of the photon sphere shrinks slowly with the growing trend of $\Lambda$ (as prescribed by the effective potential).

Consequently, we additionally used the thick disks as a probe of accretion flows around the exotic tidal wormhole solutions. Detailed discussion on this subject is therefore properly presented in the Section (\ref{sec:5}). In this section, we modify the standard thick disk \texttt{GYOTO} scenario for our needs and set the inner radius of the thick accretion disk to $r_{\mathrm{in}}=r_0$. Moreover, we keep thick disk opening angle and magnetisation parameter as defaults, therefore $\theta_{\mathrm{op}}=30^{\circ}$ and $\sigma=0.01$. Moreover, black body disk temperature and the inner radius is $T(r_{\mathrm{in}})=8\times 10^{10}$K and electron number density $n_e(r_{\mathrm{in}})=2.4\times 10^5$cm$^{-3}$, furthermore, velocity $v^\phi$ is adopted to be purely azimuthal (from that fact, if follows that accretion flow is circularly rotating). On top of the usual ray-tracing simulations we place the images, blurred up to the EHT resolution of $20\mu$as (simulations were also approximated to the EHT ones by setting the inclination angle of $i=20^\circ$ and observation band with $\nu=230$GHz). Images with both regular and blurred resolution could be seen on the Figures (\ref{fig:12}) and (\ref{fig:13}) for Schwarzschild-like and Reissner-N\"ordstrom wormholes respectively. It could be easily noticed that similar to the  accreting tori, Schwarzschild-like wormhole $r_{\mathrm{ph}}\to\infty$ as $\alpha\to\infty$ and $r_{\mathrm{ph}}\to0$ as $\Lambda\to\infty$. Keeping that fact in mind and considering that ring-like asymmetric structure were observed for M87 supermassive black hole by EHT with the radius $\approx 40\mu$as, we could constrains the values of $\alpha$ to be relatively small and positive and $\Lambda$ to be relatively big and positive in order to respect the EHT observational data.

On top of that, within this work we comprehensively investigated the nature of relativistic jet/thin accretion disk system for two wormhole models of our consideration. As it was revealed by the ray-tracing simulations of null rays, actual size of the relativistic jet for Schwarzshild-like wormhole grows with the $\alpha\to\infty$, and, in contrary, for RS wormhole relativistic jet size reduces with the $\Lambda\to\infty$. Apart from that features, we noticed that in both cases relativistic jet brightness dominates over the one for accretion disk (which could be fixed, if one will use bigger values of inner temperature $T_{\mathrm{in}}$).

The last feature, that were probed with the help of \texttt{GYOTO} open source software is the microlensing of the fixed star bulb (with varying $\phi$ coordinate) by the curvature of the various exotic wormhole models. With the help of microlensing, we were able to determine the approximate wormhole shadow radius on the Figure (\ref{fig:prelast}). As we found out, ZTF Schwarzschild wormhole shadow radius is located very closely to the wormhole throat and then it grows linearly with the $\alpha$. Alternatively, $R_{\mathrm{sh}}$ for the RS wormhole decrease with the growing $\Lambda$, as expected by the theoretical predictions.

There was some research done within the field of ray tracing for various black holes and other exotic objects. For example, in the work \citep{Vincent2021}, both Kerr and non-Kerr compact objects, such as boson stars or Lamy wormholes were probed using previously mentioned \texttt{GYOTO} code. In the aforementioned paper, for non-rotating BHs it was found that photon sphere lies inside accretion disk for $r_{\mathrm{in}}=r_{\mathrm{ISCO}}$ and outside for $r_{\mathrm{in}}=r_{\mathrm{H}}$. Since we considered the first case, our results coincide only for relatively small, positive $\alpha$ and $|\Lambda|\ll1$. On the other hand, if BH has spin (in \citep{Vincent2021}, it was assumed that BH spin equals to $a=0.8M$), even for the option with  $r_{\mathrm{in}}=r_{\mathrm{ISCO}}$, $r_{\mathrm{ph}}>r_{\mathrm{in}}$ in general. Therefore, in such case parameter space for our models is bigger, Schwarzschild-like wormhole could have tidal forces. Besides, situation with boson stars and Lamy wormholes is more complicated. On the boson star ray tracing images, no particular photon ring were observed, but EHT images of our non-tidal Schwarzschild wormhole could match the one, obtained for boson star. However, Lamy wormhole has a very non-trivial ray tracing image and several times lensed photon ring, but if we match some free parameters, low resolution EHT images will converge. In the other remarkable work, written by \citep{RAHAMAN2021115548} ray tracing images of wormholes of Teo class with spin surrounded by Keplerian disk were comprehensively probed. In that paper, numerical dependence of marginally stable orbit radius $r_{\mathrm{ms}}$ from dimensionless spin parameter $J/M^2$ were derived, while for our models, analytical solution for $r_{\mathrm{ms}}$ is localed in the Equations (\ref{eq:43}) and (\ref{eq:4444}). From that dependence, with equal throat radii one could distinguish our wormhole models from the Teo wormholes easily, which could be useful in the discussion on the nature of some compact massive objects.

Apart from the accretion disk ray tracing, torus-jet images were obtained in the paper \citep{Vincent2019} for the single case of Schwarzschild black hole. Generally, relativistic jet for our WH models does look very similar to the one, produced by the curvature of Schwarzschild BH spacetime, so only the previously discussed constraints on the accretion disk should be applied. Additionally, in this study we compared radiation flux of classical Schwarzschild black hole and WH models of our consideration. In this work, we tried to investigate the features of tidal wormholes, that could separate such WH solutions from classical black holes or other exotic objects in such observing campaings as EHT. We have not considered the accretion disk located in the other universe. We constrained parameters $\alpha$ and $\Lambda$ in order for WH solutions to match current M87 shadow observations, performed by EHT collaboration. In relation to the previous results for various other black- and wormhole models, our models could be precisely matched to the recent VLBI observations and future campaings, we could easily vary the radius of photon ring and angular size of relativistic jet. Furthermore, James Webb Space Telescope could observe lensing effects, caused by supermassive compact object. Moreover, because of the non-singular nature of such objects, many cosmological problems that arise from central singularities could be omitted. However, as it was already noticed in the many previous papers on similar topic, with the EHT resolution of 20 $\mu$as one could fine tune almost any particular model to the observational constraints and in order to better understand the nature of supermassive object being observed, one need to increase the resolution of the image.

In the next, second part of the paper series "Accretion flows around exotic tidal wormholes", we are planning to observe the magnetohydrodynamical features of the wormhole models that we studied in the present work.
\section*{Acknowledgements}
We want to thank the Thibaut Paumard for the fruitful discussion on the \texttt{GYOTO} package and the provided help, guidance. As well, we want to express our sincere gratitude to the Frederic Vincent for the assistance on the EHT synthetic image producing. The authors gratefully acknowledge the computing time provided on the high performance computing facility, Sharanga, at the Birla Institute of Technology and Science - Pilani, Hyderabad
Campus. PKS acknowledges National Board for Higher Mathematics (NBHM) under Department of Atomic Energy (DAE), Govt. of India for financial support to carry out the Research project No.: 02011/3/2022 NBHM(R.P.)/R\&D II/2152 Dt.14.02.2022. We are very much grateful to the honorable referee and to the editor for the illuminating suggestions that have significantly improved our work in terms of research quality, and presentation.  

\end{document}